\documentclass[11pt]{article}
\pdfoutput=1
\usepackage{jheppub}
\usepackage{graphicx}
\usepackage{varwidth}
\usepackage{amssymb}
\usepackage{bbm}
\usepackage{amsmath}
\usepackage{calc}
\usepackage{float}
\usepackage{slashed}
\usepackage{stmaryrd}
\usepackage{verbatim}
\usepackage{enumitem}
\usepackage{subfig}
\usepackage{booktabs}
\usepackage[table,xcdraw]{xcolor}

\newcommand\beq{\begin{equation}}
\newcommand\eeq{\end{equation}}

\newcommand{\vev}[1]{\langle #1 \rangle}

\newcommand\mc[1]{{\mathcal{#1}}}

\newcommand\CL{{\mc{L}}}
\newcommand\CM{{\mc{M}}}
\newcommand\CO{{\mc{O}}}

\title{Flavor Broken QCD$_3$ at Large N}

\preprint{\today}

\author[a]{Andrew Baumgartner \\\href{mailto:baum4157@uw.edu}{baum4157@uw.edu} }

\affiliation[a]{Department of Physics, University of Washington, \\
3910 15th Ave NE, Seattle, WA, 98195-1560, USA}

\abstract{We examine the vacuum structure of QCD$_3$ with flavor group $U(f)\times U(N_f-f)$ in the limit $N\to\infty$ with $g^2N=$fixed. We find that, generically, the resolution of critical points into a series of first order pahse transitions persists at special locations in the phase diagram. In particular, the number of Grassmannians that one traverses and their locations in the phase diagram is a function of $f$. }

\begin{document}
\maketitle

\section{Introduction}
In the past couple years a large body of work on 2+1 dimensional gauge theories has revealed a panoply of fascinating physics. Partially spurred by advances in Chern-Simons matter theories \cite{Minwalla:2015sca,Giombi:2011kc,Aharony:2011jz,Aharony:2012nh,Jain:2014nza, Jain:2013gza}, higher spin gravity \cite{Giombi:2009wh,Giombi:2012ms}, emergent gauge field \cite{Son:2015xqa} and supersymmetric dualities \cite{Kachru:2016aon,Gur-Ari:2015pca}, this research program has shed light on important concepts in particle, string and condensed matter physics (see \cite{Mross:2015idy, Mross:2017gny,Karch:2016sxi, Aitken:2017nfd, Aitken:2018joi, Kachru:2016rui, Karch:2018mer, Jensen:2017xbs, Karch:2016aux, Wang:2017txt, Chester:2017vdh, Gomis:2017ixy, Cordova:2017vab, Cordova:2017kue, Cordova:2018qvg, Benini:2017aed, Jensen:2017bjo, Chen:2018vmz, Son:2018zja, Chen:2017lkr, Armoni:2018ahv, Choi:2019eyl, Aitken:2019shs, Aitken:2019mtq, Sachdev:2018nbk} for a subset of this work). A large portion of these insights stem from the existence of the non-Abelian Chern-Simons term:
\begin{equation}\label{eq:CSterm}
S_{CS}=\frac{k}{4\pi}\int \text{Tr}\left( A\wedge dA +\frac{2i}{3}A\wedge A \wedge A\right)
\end{equation}
where the level $k$ is quantized to ensure gauge invariance and Tr refers to trace over the gauge group (we will be primarily concerned with $SU(N)$). This term generically dominates the dynamics at low energies due to the single derivative. Equation \eqref{eq:CSterm} has a long and storied history in theoretical and mathematical physics (see the review \cite{Dunne:1998qy} and the references therein), but has recently garnered attention due to its ubiquity in emergent 2+1 d gauge theories in condensed matter physics. Indeed, eq. \eqref{eq:CSterm} is unavoidable in any planar theory of fermions since fermions of mass $m$ dynamically generate a Chern-Simons term with level $\text{sgn}(m)/2$ at one loop.

An interesting story arises when one studies eq. \eqref{eq:CSterm} coupled to massless fermions. One finds the surprising result that for certain parameter regimes (namely when the level $k\ge N_f/2$ where $N_f$ is the number of fermions) \footnote{Here and throughout we use the convention of \cite{Komargodski:2017keh, Seiberg:2016gmd, Hsin:2016blu, Aharony:2016jvv, Seiberg:2016rsg} that $k=k_b-N_f/2$, where $k_b$ is the bare Chern-Simons level and $-N_f/2$ is the contribution from the $\eta$-invariant} there exists a duality \cite{Aharony:2015mjs}  : 
\begin{equation} \label{eq:aharony}
SU(N)_{k} \text{ with } N_f \, \, \psi \,\,\,\,\leftrightarrow\,\,\,\, U(\frac{N_f}{2}+k)_{-N} \text{ with } N_f \,\, \phi
\end{equation} 
where $\phi$ are Wilson-Fisher scalars with single and double trace quartic potentials \cite{Argurio:2019tvw, Armoni:2019lgb}. As one increases the number of flavors, eq. \eqref{eq:aharony} fails to hold, since the gauge group of the scalars gets completely broken down, leading to a nonlinear sigma model (NL$\sigma$M). This lead the authors of \cite{Komargodski:2017keh} to a new type of duality in which the fermions undergo spontaneous symmetry breaking for a small range masses. This scenario, valid for $k<N_f/2\le N_{\star}$, where  $N_{\star}$ is a yet-to-be pinned down upper bound, leads to the duality \footnote{Here we choose a definition of $m_{\psi}$ such that the critical points are symmetric about the origin, however this need not be the case.} 
\begin{equation}
SU(N)_{k}\,\text{with }N_{f}\,\psi \, \, \, \, \,\leftrightarrow\, \, \, \, \,\begin{cases}
U(\frac{N_f}{2}+k)_{-N}\,\text{with }N_{f}\,\phi & m_{\psi}=m_{\star}\\
U(\frac{N_f}{2}-k)_{N}\,\text{with }N_{f}\,\,\,\tilde{\phi} & m_{\psi}=-m_{\star}
\end{cases}.\label{eq:flq aharony}
\end{equation}
Between these critical points exists a NL$\sigma$M given by the complex Grassmannian:
\begin{equation}\label{eq:grassman}
Gr(\frac{N_f}{2}+k,N_f) \, \cong \,\frac{U(N_f)}{U(\frac{N_f}{2}+k)\times U(\frac{N_f}{2}-k)}
\end{equation}
supplemented by an appropriate WZW type term. This scenario assumes that each of the $N_f$ fermions is given the same mass deformation. One can relax this condition as in \cite{Baumgartner:2019frr, Argurio:2019tvw, Khalil:2020chx} and study the phase diagram as a function of distinct mass deformations for different groups of fermions. These different masses explicitly break the flavor symmetry and one obtains inequivalent Grassmannians in different part of the phase diagram. As one changes the size of these groups, these Grassmannians can disappear and reappear in at a different locations in the diagram. In addition, there exist special points in the diagram where multiple Grassmannians become degenerate. 

\begin{figure}[h] 
\centering
\subfloat[$k>N_f/2$ and $N_f\to \infty$ ]{
\includegraphics[scale=.3]{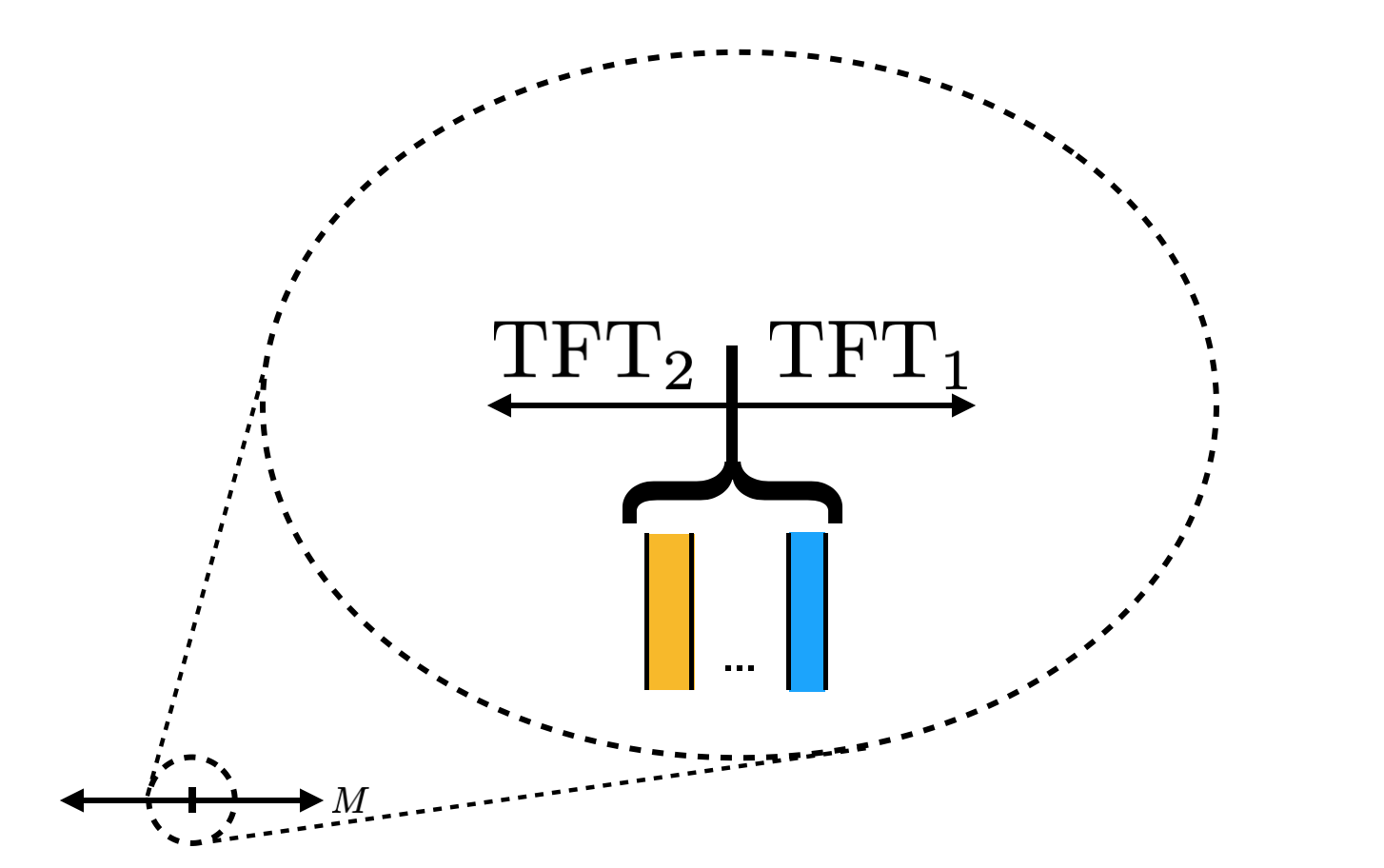}
}\subfloat[$k<N_f/2<N_{\star}$]{
\includegraphics[scale=.3]{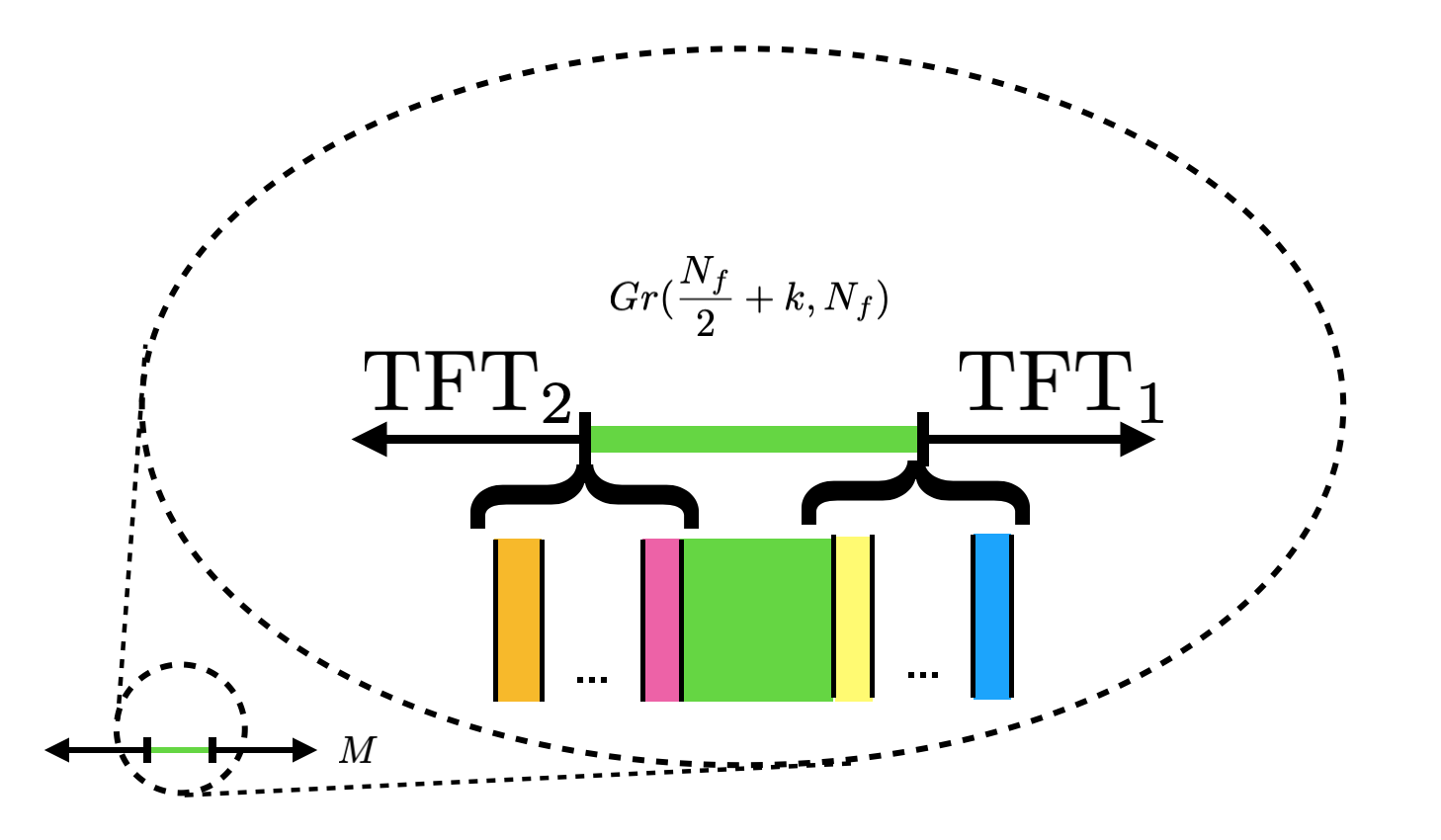}
}
\caption{Resolution of critical points into a series of first order phase transitions. The different colored regions represent distinct Grassmannians}\label{fig:resolve}
\end{figure}

The plot thickens when one passes to the large $N$ limit. Large $N$ limits in Chern-Simons matter theories have been studied before (see for instance \cite{Minwalla:2015sca,Giombi:2011kc,Aharony:2011jz,Aharony:2012nh,Jain:2014nza} and references therein) where $N/k$ remains constant. However, this limit does not lead to the Grassmannian \eqref{eq:grassman}. Instead, the authors or \cite{Armoni:2019lgb} studied a large $N$ limit of Yang-Mills-Chern-Simons theory coupled to fermions with $g^2N=const$, where $g$ is the gauge coupling. In this limit, the distinguished points in the phase diagram where one transitions between the semi-classical phases get resolved into a series of first order phase transitions with Grassmannians of the form $Gr(p,N_f)$ as in figure \ref{fig:resolve}. Interestingly, this is true when $k=0$ as well as when $k\ge N_f/2$ and $k<N_f/2$. Additionally, each Grassmannian is accompanied by a decoupled pure Chern-Simons theory of the form $SU(N)_{k+p-N_f/2}$, even when $k=0$. The phases identified at finite $N$ are distinguished by the fact that their CS theory has zero level. 

In this work, we extend the analysis of \cite{Baumgartner:2019frr, Argurio:2019tvw} to the large $N$ limit using the techniques laid out in \cite{Armoni:2019lgb}. We find series of first order phase transitions in special locations of the phase diagram consistent with the results of \cite{Baumgartner:2019frr, Argurio:2019tvw} . Interestingly, there exists special places in the diagram where many ($2n$ where $n-1$ is the number of distinct flavors) Grassmannians become degenerate. Additionally, for odd $N_f$ we find degenerate symmetry broken and symmetry unbroken phases from any value of the mass, not just at special locations where phase transitions occur. We also discuss modifications to the dual scalar potential which will lead to doubly symmetry broken phases. The layout of our paper is as follows: in section \ref{sec:2} we review the necessary results of \cite{Komargodski:2017keh,Armoni:2019lgb,Baumgartner:2019frr, Argurio:2019tvw} which are used in this work. Next, in section \ref{eq:meat}, we discuss the construction of our phase diagram and perform some consistency checks. In section \ref{scalars} we briefly discuss scalar potentials and modifications which lead to interesting doubly symmetry broken phases. We conclude in section \ref{sec:con}. Additionally, we include numerical evidence for the diagram in appendix \ref{sec:append}. 

\section{Symmetry Breaking in QCD$_3$ and Vacuum Structure at Large N}\label{sec:2}
As mentioned in the introduction, the 3d non-Abelian bosonization dualities fail to hold when $k\le N_f/2<N_{\star}$ where $N_{\star}$ is some yet-to-be-pinned-down function of $N_f$ and $k$ \footnote{For an upper bound derived from the f-theorem see \cite{Sharon:2018apk}}. When the scalars obtain a large negative mass deformation, they obtain a vev of the form 
\begin{equation}
\langle \phi \rangle = \left( \mathbbm{1}_{\frac{N_f}{2}+k}\, \, 0_{\frac{N_f}{2}-k} \right)
\end{equation}
which spontaneously breaks the flavor symmetry as 
\begin{equation}
U(N_f)\to U(\frac{N_f}{2} + k)\times U(\frac{N_f}{2}-k )
\end{equation}
leading to Goldstone bosons which are valued in the complex Grassmannian \footnote{Henceforth we will drop the ``complex" for sake of presentation.}
\begin{equation}\label{eq:GrassmanDef}
Gr(N_f/2+k,N_f) \, \cong \, \frac{U(N_f)}{U(\frac{N_f}{2} + k)\times U(\frac{N_f}{2}-k )} .
\end{equation}
A key observation of \cite{Komargodski:2017keh} is that this is not the only scalar theory with this symmetry breaking pattern. One can instead consider the theory 
\begin{equation}
U(\frac{N_f}{2}-k)_{-N} \text{ with } N_f \,\, \tilde{\phi}
\end{equation}
with a large negative mass deformation and find the same symmetry breaking pattern, leading to equivalent low energy dynamics. In addition, positive mass deformations of this scalar theory will land us in the level/rank dual of the TFT corresponding to negative mass deformations of the original fermionic theory. This motivates the duality 
\begin{equation}\label{eq:duality}
SU(N)_{k}\,\text{with }N_{f}\,\psi \, \, \, \, \,\leftrightarrow\, \, \, \, \,\begin{cases}
U(\frac{N_f}{2}+k)_{-N}\,\text{with }N_{f}\,\phi & m_{\psi}=m_{\star}\\
U(\frac{N_f}{2}-k)_{N}\,\text{with }N_{f}\,\,\,\tilde{\phi} & m_{\psi}=-m_{\star}
\end{cases}
\end{equation}
and the corresponding phase diagram in figure \ref{fig:resolve}. Thus, there is a finite range of masses for which spontaneous symmetry breaking (SSB) occurs and the low energy theory is described by massless Goldstone modes valued in the Grassmannian. In addition to the Lagrangian which describes the NL$\sigma$M, the action contains a contribution from a Wess-Zumino-Witten term which allows for the existence of skyrmions which play the role of baryons in the symmetry broken phase. The WZW term, however, will not play a crucial role in our discussion. 

\subsection{Symmetry Breaking with Unequal Masses}\label{unequal}
\begin{figure}[h] 
\centering
\subfloat[Phase diagram associated to i).]{
\includegraphics[scale=.4]{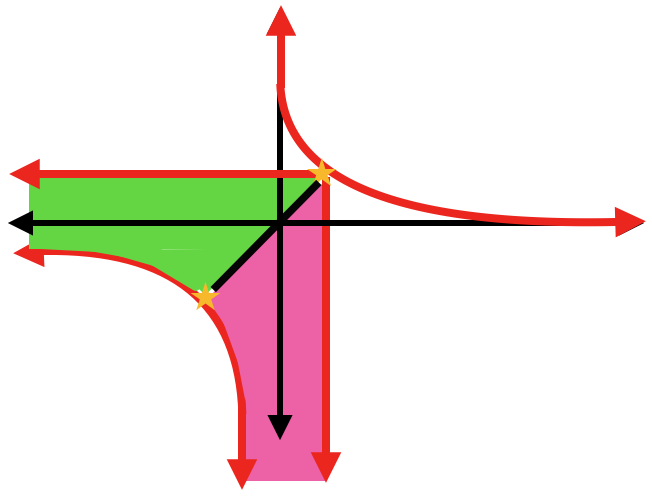}
}\hfill
\subfloat[Phase diagram associated to ii).]{
\includegraphics[scale=.4]{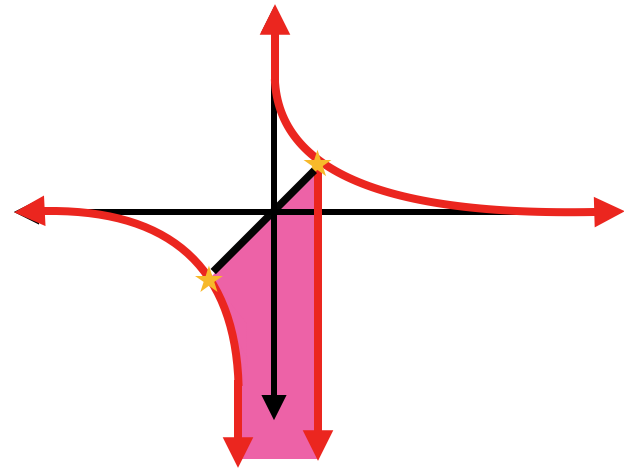}
}\hfill
\subfloat[Phase diagram associated to iii).]{
\includegraphics[scale=.4]{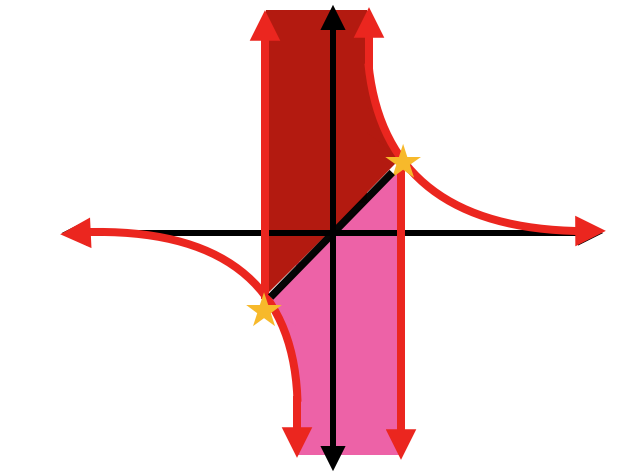}
}\vfill

\subfloat[Phase diagram associated to iv).]{
\includegraphics[scale=.4]{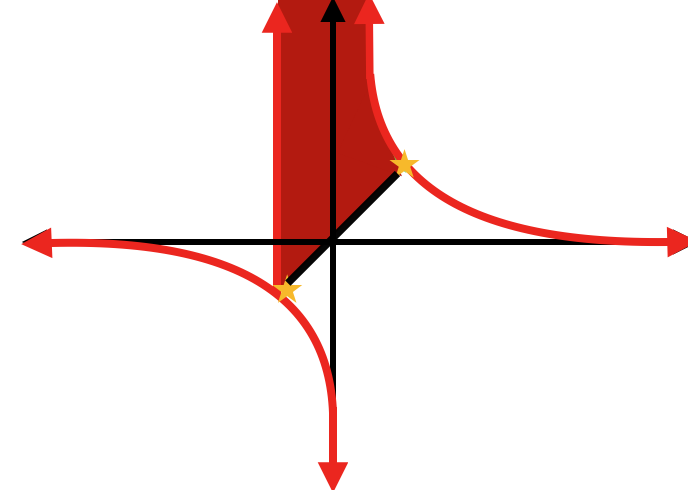}
} \hspace{.7 in}
\subfloat[Phase diagram associated to v).]{
\includegraphics[scale=.4]{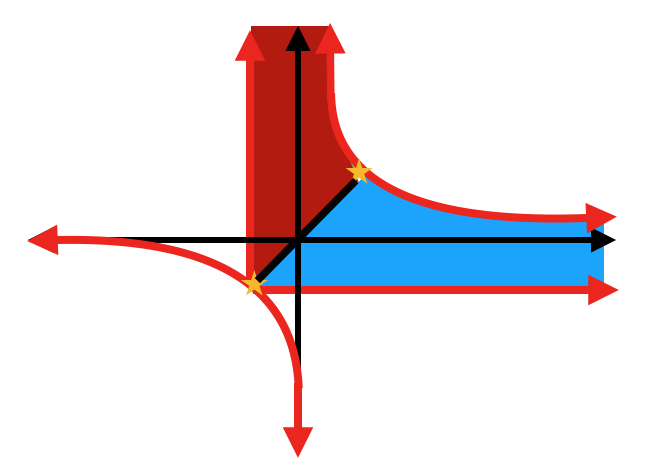}
}
\caption{Various phase diagrams for the scenarios discussed in sec. \ref{unequal}, where $m$ is on the $y$ axis and $M$ is on the x-axis.}\label{uneqDiags}
\end{figure}

The above analysis is applicable when one gives equal masses to all $N_f$ of the fermions. A more interesting structure emerges when one gives different mass deformations to separate subsets of the $N_f$ fermions \cite{Baumgartner:2019frr, Argurio:2019tvw}. Namely, we give the first $f$ of the fermion flavors a mass $m$ and the other $N_f-f$ a mass $M$. Such mass deformations will explicitly break the flavor group as $U(N_f)\to U(f)\times U(N_f-f)$. As a result of this explicit breaking, it will always be either one or the other subset of flavor which will condense to form Grassmannians. The specific Grassmannians and their location in the phase diagram depends on the relative values of $f, N_f-f$ and $k$. There are 6 possible options:
\begin{enumerate}[label=\roman*] 
\centering
 \item[i.)] $f-\frac{N_f}{2}<k$, $\frac{N_f}{2}-f<k$
 \item[ii.)] $f-\frac{N_f}{2}<k$, $\frac{N_f}{2}-f=k$
 \item[iii.)] $f-\frac{N_f}{2}<k$, $\frac{N_f}{2}-f>k$
 \item[iv.)] $f-\frac{N_f}{2}=k$, $\frac{N_f}{2}-f>k$
\item[v.)] $f-\frac{N_f}{2}>k$, $\frac{N_f}{2}-f>k$
\item[vi.)] $f-\frac{N_f}{2}=k$,$\frac{N_f}{2}-f=k$.
\end{enumerate}
The structure of the phase diagrams in each case shown in figure \ref{uneqDiags}. For each of parameter regime we have the following Grassmannians \footnote{These are in addition to the Grassmannian $Gr(k+\frac{N_f}{2},N_f)$ which exists along the diagonal in each regime} along the specified axis in the $m-M$ plane. 
\begin{center}
\begin{varwidth}{\textwidth}
\begin{enumerate}[label=\roman*] 
 \item[i.)] $M<0: \, Gr(k+f-\frac{N_f}{2},f)$ \\
 \qquad  $m<0: \, Gr(k-f+\frac{N_f}{2},N_f-f)$
 \item[ii.)] $m<0: \, Gr(k-f+\frac{N_f}{2},N_f-f)$ 
 \item[iii.)] $m>0: \, Gr(k+\frac{N_f}{2},N_f-f)$ \\
 \qquad $m<0: \,  Gr(k-f+\frac{N_f}{2},N_f-f)$ 
 \item[iv.)] $m>0: \, Gr(k+\frac{N_f}{2},N_f-f)$ 
\item[v.)] $M>0: \, Gr(k+\frac{N_f}{2},f)$\\
\qquad $m>0: \, Gr(k+\frac{N_f}{2},N_f-f)$
\item[vi.)] No additional Grassmannians
\end{enumerate}
\end{varwidth}
\end{center}
Strictly speaking this analysis is applicable when at least one mass deformation is macroscopically large \footnote{By ``macroscopically large" we mean larger than the strong scale $\Lambda$.}. The massive fermions shift the Chern-Simons level and the problem then reduces to the analysis of flavor bounds of the light flavors with this shifted level. The behavior when both mass deformations are small is still ambiguous but can be slightly elucidated in the large $N$ limit. 

\subsection{Vacuum Structure at Large N}
One major drawback of the analyses in the previous section is the lack of insight about the nature of the phase transitions. On one hand the fact that they are mediated by scalar theories seems to indicate that these are second order. However they are also strongly coupled and so the presence of light degrees of freedom at the transition point does not necessarily mean that the transition is second order. In order to probe the nature of these phase transitions, we must pass to some perturbative limit where calculations are possible. 

One such perturbative analysis was recently performed in \cite{Armoni:2019lgb} using the ``standard" 't Hooft large 
$N$ limit: $N\to \infty, \, g^2N=$fixed. This differs from previous large $N$ calculations in Chern-Simons matter theories where one takes $k/N=$ fixed. As a result, this counting scheme leads to the standard large $N$ counting rules--leading order diagrams are arbitrary planar gluon diagrams, and quark loops are suppressed by a factor of $1/N$. 

Since we are interested in the patterns of flavor-symmetry breaking in this limit, the primary observable of interest is the expectation value of the fermion condensate
\begin{equation}
\CM_I^{\,J}=\frac{1}{N} \langle \bar{\psi}_I \psi^{J} \rangle
\end{equation}
where $I,J=1,..,N_f$ are flavor indices. This term is valued in the adjoint representation of the global $SU(N_f)\subset U(N_f)$ flavor symmetry, invariant under the baryonic $U(1)\subset U(N_f)$ and is time-reversal odd. It enters into the Lagrangian via a mass term for the fermions
\begin{equation}
\CL \supset -m_I^{\,J} \bar{\psi}_J \psi^{I}=-N \text{tr}(m\CM).
\end{equation}
Here we defined $m$ as the matrix with entries $m_I^{\,J}$.

The basic strategy of \cite{Armoni:2019lgb} is to use large $N$ reasoning to derive a general form for the effective potential as a function of $\langle \CM \rangle $. Specifically, we are interested in 
\begin{equation}\label{eq:pot_def}
V(\langle \CM \rangle ) = -W(m)-N \text{tr}(m\CM)
\end{equation}
where $W(m)=-i\text{ln}(Z(m))/V_3$ is the sum of all connected correlation functions of $\CM$ and $V_3$ is the spacetime volume. We now review their leading order and next to leading order calculations.
\subsubsection*{Leading Order}
\begin{figure}[h] 
\centering
\subfloat[ ]{
\includegraphics[scale=.5]{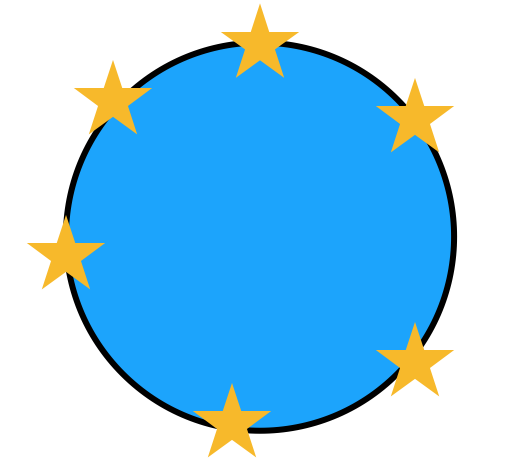}
} \hspace{.5 in} \subfloat[ ]{
\includegraphics[scale=.5]{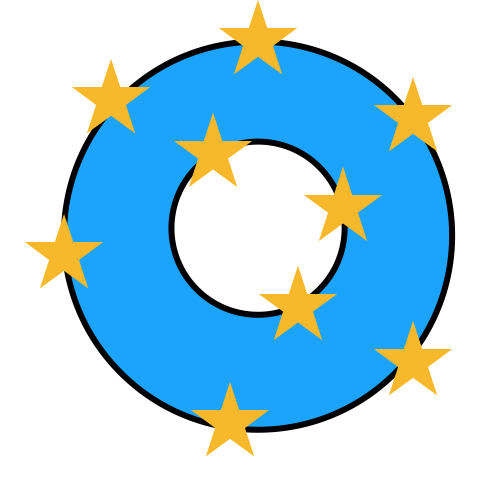}
}
\caption{Leading (a) and subleading (b) diagrams in the large N limit. Blue represents arbitrary gluon loops and stars are insertions of $\mathcal{M}$.}\label{fig:diskannul}
\end{figure}

The $\CO(N)$ contributions to $W(m)$ (and, by extension, $V(\langle \CM \rangle)$ ) come from diagrams which are topologically a disk with a single fermion loop and an arbitrary number of $\CM$'s inserted on it's boundary as in figure \ref{fig:diskannul}. The interior of the disk is an arbitrary planar gluon diagram. As a result, $W(m)$ can only take on contributions from single-trace operators. Time reversal symmetry requires that the potential obey $W(-m)=W(m)$ which further constrains the allowed terms. When the dust settles, we can schematically write $W(m)$ as 
\begin{equation}\label{eq:connectDiag}
W(m)=N\Lambda^3\sum_{n=2,4,6..}^{\infty} \frac{C_n}{\Lambda^n}\text{tr}(m^n)
\end{equation}
which leads to a very similar form for the effective potential:
\begin{equation}\label{eq:eff_Pot_full}
V(\langle \CM \rangle)=N\Lambda^3\sum_{n=2,4,6..}^{\infty} \frac{C'_n}{\Lambda^{2n}}\text{tr}(\langle \CM \rangle^n).
\end{equation}
We now use an $SU(N_f)$ transformation to diagonalize $\CM$ to get 
\begin{equation}\label{eq:ferm_cond}
\CM=\Lambda^3\text{diag}(x_1,x_2...x_{N_f}),\, \, \, \, x_i \in \mathbb{R}
\end{equation}
which leads to an effective potential of the form 
\begin{equation}\label{eq:eff_pot_LO}
V(x_i)=N\Lambda^3\sum_{i=1}^{N_f}\sum_{n=2,4,6,...}^{\infty} C'_nx_i^n\equiv N\Lambda^3\sum_{i=1}^{N_f}F(x_i).
\end{equation}
An incredibly crucial assumption of \cite{Armoni:2019lgb} is that $F(x)$ has degenerate minima at some finite $\pm x_{\star}$ and by a simultaneous rescaling of $\CM$ and $F(x)$ we can set $x_{\star}=1$ \footnote{Without this assumption the results of \cite{Armoni:2019lgb} would be inconsistent with the Vafa-Witten theorem \cite{Vafa:1984xh, Vafa:1983tf}}. Since the potential is minimized as $x_i=\pm 1$, each eigenvalue must take on either one of those values. This leads to $N_f+1$ degenerate vacua labeled by the number of positive (or negative) eigenvalues of $\CM$. The NL$\sigma$M that describes the low energy physics is again the Grassmannian
\begin{equation}\label{eq:Grassmann}
Gr(p,N_f)\cong \frac{U(N_f)}{U(p)\times U(N_f-p)}
\end{equation}
supplemented by the appropriate WZW term. In addition to the Grassmannian, each of these vacua will contain a non-trivial Chern-Simons TFT of the form $SU(N)_{p-\frac{N_f}{2}+k}$, even when $k=0$. This is necessary is one assumes that there are no phase transitions as one tunes the mass from small non-zero values to asymptotically large values. 

One can add a mass term for the fermions which will lift the degeneracy of these vacua. It is not difficult to show that 
\begin{equation}
\left[ m,\CM \right]=0
\end{equation}
which allows for the simultaneous diagonalization of both $m$ and $\CM$. Let $m_i$ be the eigenvalues of $m$. Then the mass term will add a term to the effective potential of the form
\begin{equation}\label{eq:pot_NLO}
V(x_i)\supset N\Lambda \sum_{i=1}^{N_f} m_i x_i.
\end{equation}
Now if we take the first $m_i>0$ for $i=1,..,p$ and $m_i<0$ for $i=p+1,...,N_f$, the potential will be minimized by taking $x_i=-1$ for $i=1,..,p$ and $x_i=1$ for $i=p+1,...,N_f$. This lifts the degeneracy and singles out $Gr(p,N_f)$ as the true ground state. 

\subsubsection*{Next-To-Leading-Order}
In the proceeding section we discussed the vacuum structure of QCD$_3$ by examining the $\CO(N)$ contribution to the effective potential, which stems from the disk diagram of figure \ref{fig:diskannul} (a). The $\CO(1)$ contribution, on the other hand, comes from the annulus diagram of figure \ref{fig:diskannul} (b) and so contains double trace terms such as $\text{tr}(\langle \CM \rangle ^n)\text{tr}(\langle \CM \rangle ^{n'})$. Since we are not interested in any higher order terms in the large N expansion, we can restrict the $x_i$'s to their values which minimize the $\CO(N)$ potential. This fact, along with constraints from time-reversal invariance, give the following form for the $\CO(1)$ potential
\begin{equation}\label{eq:Gterm}
V(x_i)\supset \Lambda^3\Delta \sum_{i,j=1}^{N}x_ix_j.
\end{equation}
The constant $\Delta$ is a strictly positive constant whose value is determined from calculation of the diagrams in figure \ref{fig:diskannul}. It's positivity is required by consistency with the Vafa-Witten theorem--in the absence of mass terms the ground state is given by $Gr(N_f/2,N_f)$. 

One can add a small singlet mass deformations $m \sim m\delta_{ij}$ to this and examine how the ground state will change as a function of the mass. In addition, we can add a non-zero Chern-Simons term to the action. The contribution from the Chern-Simons term enters in the same form as the mass term and so we can group them together. The potential is then 
\begin{equation}\label{eq:effPotUnbrokeLO}
V(x_i)\supset \Lambda^3\Delta \sum_{i,j=1}^{N_f}x_i x_j +\left(Nm \Lambda^2+k\Lambda^3 \right)\sum_{i=1}^{N_f} x_i.
\end{equation}
Note that this mass deformation is subleading only when $m~\CO(1/N)$. If $m \sim \CO(1)$, then it becomes part of the $\CO(N)$ potential. Now we make the ansatz that the first $p$ eigenvalues of $\langle \CM \rangle$ are -1 and the remaining $N_f-p$ are 1. Minimizing the effective potential with respect to $p$ gives 
\begin{equation}\label{eq:min_p_unbroke}
p=\frac{N_f}{2}+\llbracket \frac{Nm+\Lambda k}{4\Lambda \Delta} \rrbracket
\end{equation}
where $\llbracket x \rrbracket$ denotes the nearest integer of $x$. The vacuum which minimizes the effective potential jumps by $\pm 1$ whenever $ \frac{Nm+\Lambda k}{4\Lambda \Delta}  \in \mathbb{Z} +\frac{1}{2}$. Thus as we tune $m$ we traverse each Grassmannian until we wind up in the asymptotic phases. In other words, one encounters a series of first order phase transitions as a function of $m$.

\section{Vacuum Structure With an Explicitly Broken Flavor Group}\label{eq:meat}
In the above discussion we gave the fermions a mass deformation of the form $m=m\mathbbm{1}$ which preserved the global flavor symmetry. In this section we will examine a different scenario in which
\begin{equation}\label{eq:massDefBroke}
\tilde{m}=\text{diag}\left(\underbrace{m,..,m}_{f},\underbrace{M,..,M}_{N_{f}-f}\right).
\end{equation} \label{eq:ferm_cond_broke}
This mass deformation \textit{explicitly} breaks the global flavor symmetry as $U(N_f)\to U(f)\times U(N_f-f)$. As a result, we will distinguish the corresponding eigenvalues of our condensate: 
\begin{equation}
\langle \CM \rangle = \frac{1}{N}\text{diag}\left(x_{1},..,x_{f},y_{1},...,y_{N_{f}-f}\right).
\end{equation}
We may then ask if these subgroups get broken down any further by the resulting fermion condensate. The purpose of this section is to determine the circumstances under which these subgroups get spontaneously broken. 

\subsection{Leading Order}
We start by assuming $M=0$. The full $\CO(N)$ potential including the mass deformation is
\begin{equation} \label{eq:eff_pot_broke_LO}
V(x_{i})=N\Lambda^{3}\sum_{i=1}^{N_{f}}F(x_{i})+m\sum_{i=1}^{f}x_{i}.
\end{equation}
The mass term fixes all the $x$'s to be $-1$ but places no such restriction on the $y's$. Thus each $y$ can be $\pm 1$ which again leads to a degeneracy. Assuming the first $p$ of the $y$ eigenvalues are $1$, our symmetry group will spontaneously break as
\begin{equation}\label{eq:SSBwEFB}
U(f)\times U(N_f-f)\to U(f)\times U(p) \times U(N_f-f-p)
\end{equation}
which leads to the coset
\begin{equation}
\frac{ U(f)\times U(N_f-f)}{U(f)\times U(p) \times U(N_f-f-p)}=\frac{ U(N_f-f)}{ U(p) \times U(N_f-f-p)} \cong Gr(p,N_f-f).
\end{equation}
Thus there are $N_f-f+1$ degenerate vacua, one for each choice of $p$, each one accompanied by a Chern-Simons TFT. 

If we take $M\ne 0$, then the corresponding eigenvalues will also be fixed and, in general, no SSB will occur. 

\subsection{Next-To-Leading Order}
The story becomes much more interesting at $\CO(1)$ due to the asymmetry in the mass term. The full $\CO(1)$ potential including, including the Chern-Simons term and an $\CO(1/N)$ mass deformation, can be written as
\begin{equation} \label{eq:Ord1Pot}
V=\Delta\Lambda^3 \left(\sum_{i=1}^{f}x_i + \sum_{i=1}^{N_f-f}y_i \right)^2+k\Lambda^3\left(\sum_{i=1}^{f}x_i + \sum_{i=1}^{N_f-f}y_i\right) + Nm\Lambda^2\sum_{i=1}^{f}x_i+ NM\Lambda^2\sum_{i=1}^{f}y_i.
\end{equation}
As in the flavor symmetric case the eigenvalues of $\CM$ take on their $\CO(N)$ values in the $\CO(1)$ potential since we are not considering terms which are higher order. The general structure of the phase diagram will also be uneffected by the Chern-Simons term and so the bulk of our analysis will be concerned with the $k=0$ case. Adding non-zero $k$ will only serve to shift the locations of various phases in the phase diagram, rather than change it's structure. 

We start this section by demonstrating a lack of double condensation of the fermions when $m\ne M$. That is, it is forbidden to have SSB in both groups of fermions simultaneously. The resulting scenarios can then be broken up into three cases--one where symmetry breaking happens in the upper left $f\times f$ block of $\CM$, one that happens in the lower right $(N_f-f)\times (N_f-f)$ block, and one where no symmetry breaking occurs. We then identify the regions of the phase diagram where these symmetry breaking scenarios occur by examining the relative values of the effective potential. We then use the fact that $Gr(M,N)$ does not exist unless $0\le M \le N$ to further refine our diagram. 

\subsubsection{A Lack of Double Condensation}
We start by looking for a solution in which both groups of fermions undergoes spontaneous symmetry breaking. That implies the formation of a condensate of the form 
\begin{equation} \label{eq:doubleCon}
N\vev{\CM}=\text{diag}(\underbrace{1,...,1}_p,\underbrace{-1,...,-1}_{f-p},\underbrace{1,...,1}_q\underbrace{,-1...,-1}_{N_f-f-q})
\end{equation}
leading to the double coset
\begin{equation}\label{eq:doubleCos}
Gr(p,f)\times Gr(q,N_f-f).
\end{equation}
The integers $p$ and $q$ are bounded below by zero and above by the number of fermion flavors in that block. In other words, $p$ and $q$ takes integer values in the domain $\mc{D}=[0,f]\times[0,N_f-f]$. If, for general $m$ and $M$, the effective potential admits a minimum on $\text{int}(\mc{D})$ then double condensation can occur. Otherwise, the minimum must occur on $\partial \mc{D}$ and only one group of flavors undergoes SSB. Plugging \eqref{eq:doubleCon} into \eqref{eq:Ord1Pot} gives 
\begin{equation}\label{eq:doublePot}
\frac{V}{\Delta\Lambda^3}=\left(2p+2q-N_f \right)^2+\left(\frac{mN}{\Lambda \Delta}+\frac{k}{\Delta} \right) \left(2p-f\right) +\left(\frac{MN}{\Lambda \Delta}+\frac{k}{\Delta} \right)\left(2q-N_f+f\right).
\end{equation}
To find extrema, we extend $p$ and $q$ to the real numbers and solve the system of equations:
\begin{align}\label{eq:sysEq}
\frac{\partial V}{\partial p}&=4(2p+2q-N_f)+2\left(\frac{mN}{\Lambda \Delta}+\frac{k}{\Delta} \right)=0 \\
\frac{\partial V}{\partial q}&=4(2p+2q-N_f)+2\left(\frac{MN}{\Lambda \Delta}+\frac{k}{\Delta} \right)=0.
\end{align}
To simplify this, rewrite it as a matrix equation:
\begin{equation}
\left( \begin{array}{cc}
8 & 8\\
8 & 8
\end{array}\right)\left( \begin{array}{c} p \\ q \end{array}\right)=\left(\begin{array}{c} 4N_f -2\left(\frac{mN}{\Lambda \Delta}+\frac{k}{\Delta} \right) \\ 4N_f -2\left(\frac{MN}{\Lambda \Delta}+\frac{k}{\Delta} \right) \end{array}\right). 
\end{equation}
Clearly the coefficient matrix $\left( \begin{array}{cc}
8 & 8\\
8 & 8
\end{array}\right)$ has determinant zero and is therefore not invertible. This system of equations has no solutions and the extrema must occur on $\partial \mc{D}$.  Explicit numerics demonstrating the same conclusion are given in Appendix A. 

At finite $N$, one can show that double condensation is forbidden by explicitly minimizing the potential on the scalar side, as was done in \cite{Argurio:2019tvw, Armoni:2019lgb}. There the authors included both a single and double flavor trace quartic potentials and minimized with respect to the eigenvalues. Positivity constraints show that there is no double condensation outside of the flavor enhanced $m=M$ line. However, this same reasoning can not be applied to large $N$ QCD$_3$, since the proposed form of the (finely tuned) \textit{sextic} potential on the scalar side does not include double trace deformations at leading order. Since the leading order solution determines the eigenvalues of the condensate, there is no \textit{a priori} reason to expect a lack of double condensation. We will discuss some modifications to the potential that could allow for such double condensation in section \ref{scalars} below. 

\subsubsection{Single Massive Flavor}
We start by examining what happens when we give the first $f$ flavors a mass $m>0$ while leaving the other $N_f-f$ light. The situation is slightly different when $f\le\frac{N_f}{2}$ and $f \ge \frac{N_f}{2}$. We examine these in turn. The potential is the same in either case. We start by taking $k=0$.  We have
\begin{equation} \label{eq:Ord1Pot2}
V=\Delta\Lambda^3 \left(\sum_{i=1}^{f}x_i + \sum_{i=1}^{N_f-f}y_i \right)^2+ Nm\Lambda^2\sum_{i=1}^{f}x_i.
\end{equation}
First, consider giving $f<\frac{N_f}{2}$ a mass $m$. In this scenario, we can take all the $x$'s to be $-1$. Now, since there are more $y's$ than there are $x$'s, we can entirely cancel off the contribution from the $x$'s in the quadratic term by taking $f$ of the $y's$ to be equal to $+1$. Of the remaining $N_f-2f$ $y$'s, we take half to be positive and half to be negative so that they cancel among themselves. If $N_f$ is even, this works perfectly. If $N_f$ is odd, then we will be left with one unpaired eigenvalue. This eigenvalue can be either $+1$ or $-1$ and since it's only contribution to the effective potential is in the quadratic term, these scenarios have the same value of the effective potential and so are degenerate. When the dust settles our condensate looks like
\begin{align} \label{eq:firstScenario}
\begin{split}
N\CM&=\text{diag}\left(\underbrace{-1,...-1}_{f},\underbrace{1,...,1}_{\frac{N_{f}}{2}}\underbrace{-1,...,-1}_{\frac{N_f}{2}-f}\right)\,\,\,\,\text{ for }N_f \text{ even} \\
&=\text{diag}\left(\underbrace{-1,...-1}_{f},\underbrace{1,...,1}_{\frac{N_{f}\pm1}{2}}\underbrace{-1,...,-1}_{\frac{N_f\mp1}{2}-f}\right)\,\,\,\,\text{ for }N_f \text{ odd} 
\end{split}
\end{align} 
which leads to the Grassmannians
\begin{align}
\begin{split} \label{eq:GrassmansFirstScenario}
Gr(\frac{N_f}{2},N_f-f) & \cong Gr(\frac{N_f}{2}-f,N_f-f)\,\,\,\,\text{ for }N_f \text{ even} \\
Gr(\frac{N_f\pm1}{2},N_f-f)&\cong Gr(\frac{N_f\mp 1}{2}-f,N_f-f)\,\,\,\,\text{ for }N_f \text{ odd}
\end{split}
\end{align} 
with effective potentials given by 
\begin{align}
\begin{split}
V&=-Nm\Lambda^2f,\quad\text{ for }N_{f}\text{ even }\\
&=1-Nm\Lambda^2f,\quad\text{ for }N_{f}\text{ odd}.
\end{split}
\end{align}
These only occur, however, when 
\begin{equation}\label{eq:restrict}
\begin{split}
\frac{N_f}{2}-f\ge 1 \quad\text{ for }N_{f}\text{ even }\\
\frac{N_f \mp 1}{2}-f \ge 1\quad\text{ for }N_{f}\text{ odd}.
\end{split}
\end{equation}
Otherwise, the Grassmannians \eqref{eq:GrassmansFirstScenario} do not exist. If \eqref{eq:restrict} is not satisfied, then we simply have some Chern-Simons matter theory with $N_f-f$ massless fermions.

When $f=\frac{N_f}{2}$, we get no spontaneous symmetry breaking when $N_f$ is even. This is because all of the $y$'s can perfectly cancel all the $x$'s. We wind up with a condensate of the form:
\begin{equation}\label{eq:Nfeven}
N\langle \mc{M} \rangle=\text{diag}\left(\underbrace{-1,...-1}_{N_{f}/2},\underbrace{1,...,1}_{N_{f}/2}\right).
\end{equation}
When $N_f$ is odd, we have two choices, since not all of the $y$'s can cancel off the $x$'s. We are left with one unpaired $y$ eigenvalue which can take either sign. Since this eigenvalue enters in eq. \eqref{eq:Ord1Pot2} in the quadratic term, both choices have the same value of the effective potential and we again have a twofold degeneracy as in eq. \eqref{eq:GrassmansFirstScenario}. This leads to the following condensates: 
\begin{align}
\begin{split}
N\langle \mc{M} \rangle &=	\text{diag}\left(\underbrace{-1,...-1}_{(N_{f}-1)/2},\underbrace{1,...,1}_{(N_{f}+1)/2}\right)  \implies\text{no SSB} \\ 
	N\langle \mc{M} \rangle &=	\text{diag}\left(\underbrace{-1,...-1}_{(N_{f}-1)/2},\underbrace{1,...,1}_{(N_{f}-1)/2},-1\right) \implies\text{SSB with }Gr\left(1,\frac{N_{f}+1}{2}\right).
\end{split}
\end{align}

Already we can see some novelties of the flavor broken case as opposed to the flavor symmetric one. Namely that one does not encounter a series of first order phase transitions as you dial $m$--you simply stay in the same Grassmannian until you reach the asymptotic regime. Also we see that for odd $N_f$ there exists two degenerate Grassmannians for any value of $m$. When $f=(N_f-1)/2$, this degeneracy is between a phase where SSB occurs and one where it does not.

When $f>N_f/2$ the $y$'s can not fully cancel off all of the $x$'s. As a result, we will have some excess energy in either the quadratic term or the linear term. To proceed we calculate the extrema of the effective potential in both cases and compare the values of $V$ to see which scenario is preferred. Proceeding in this way, we find that having excess in the linear term is energetically favorable. This leads to a condensate of the form 
\begin{equation}\label{eq:fBiggerCond}
N\langle \mc{M} \rangle =	\text{diag}\left( \underbrace{-1,...-1}_{f-q},\underbrace{1,...,1}_{q} \underbrace{1,1,...1}_{N_f-f}\right) 
\end{equation}
where 
\begin{equation}
q=f-\frac{N_{f}}{2}-\llbracket \frac{Nm}{4\Lambda\Delta}\rrbracket.
\end{equation}
In contrast to the $f\le N_f/2$ case, we do encounter a series of phase transitions as we tune the mass, albeit a reduced number of them.

\subsubsection{Double Massive Flavors}

The diagram become much richer when we give the other set of flavor a mass.  The potential is 
\begin{equation} \label{eq:Ord1Pot3}
V=\Delta\Lambda^3 \left(\sum_{i=1}^{f}x_i + \sum_{i=1}^{N_f-f}y_i \right)^2+ Nm\Lambda^2\sum_{i=1}^{f}x_i+ NM\Lambda^2\sum_{i=1}^{N_f-f}y_i.
\end{equation}
Now, we must carefully examine the values of the effective potential on the boundary of our $p-q$ domain. The full domain is given by 
\begin{equation}\label{eq:pqDom}
\mc{D}=[0,f]\times[0,N_f-f]
\end{equation}
with boundary $\partial \mc{D}=\partial \mc{D}_1 \cup \partial \mc{D}_2
\cup \partial \mc{D}_3 \cup \partial \mc{D}_4$ where

\begin{align}
\begin{split}
\partial \mc{D}_1 &=[0,f]\times \{ 0\} \\
\partial \mc{D}_2 &=[0,f]\times \{N_f-f\}\\
\partial \mc{D}_3 &=\{0\}\times [0,N_f-f]\\
\partial \mc{D}_4 &=\{f\}\times [0,N_f-f]
\end{split}
\end{align} 
Let $\tilde{m}=\frac{mN}{\Lambda \Delta}$ and similarly for $\tilde{M}$. The minimum occur at:
\begin{align} \label{eq:scenarios}
\begin{split}
\partial \mc{D}_1:\,  p&=\frac{N_f}{2} -\llbracket \frac{\tilde{m}}{4} \rrbracket,\,q=0 \,\, \text{with} \,\, V_1= -\frac{\tilde{m}^2}{4}+(\tilde{m}-\tilde{M})(N_f-f)\\
\partial \mc{D}_2 : p&=f-\frac{N_f}{2} -\llbracket \frac{\tilde{m}}{4} \rrbracket,\, q=N_f-f\,\, \text{with} \,\, V_2= -\frac{\tilde{m}^2}{4}-(\tilde{m}-\tilde{M})(N_f-f)\\
\partial \mc{D}_3 : p&=0,\, q=\frac{N_f}{2} -\llbracket \frac{\tilde{M}}{4} \rrbracket\,\, \text{with} \,\, V_3= -\frac{\tilde{M}^2}{4}-(\tilde{m}-\tilde{M})f\\
\partial \mc{D}_4 : p&=f,\, q=\frac{N_f}{2}-f -\llbracket \frac{\tilde{M}}{4} \rrbracket\,\, \text{with} \,\, V_4= -\frac{\tilde{M}^2}{4}+(\tilde{m}-\tilde{M})f\\
\end{split}
\end{align} 
where $p$ and $q$ are subject to
\begin{equation} \label{eq:constraint}
0 \le p \le f,\,  \text{ and } \,0\le q \le N_f-f. 
\end{equation}
Plugging eq. \eqref{eq:scenarios} into eq. \eqref{eq:constraint} gives the following mass ranges for which we remain on $\partial \mc{D}$: 
\begin{align} \label{eq:ranges}
\begin{split}
\partial \mc{D}_1: &\,  \frac{\tilde{m}}{4} \in \mc{I}_1 \equiv \left[\frac{N_f}{2}-f , \frac{N_f}{2} \right] \\
\partial \mc{D}_2 : &\, \frac{\tilde{m}}{4} \in \mc{I}_2 \equiv \left[ -\frac{N_f}{2}, f-\frac{N_f}{2} \right]  \\ 
\partial \mc{D}_3 : &\,  \frac{\tilde{M}}{4} \in \mc{I}_3 \equiv \left[ f-\frac{N_f}{2}, \frac{N_f}{2}  \right] \\
\partial \mc{D}_4 : &\,  \frac{\tilde{M}}{4} \in \mc{I}_4 \equiv \left[ -\frac{N_f}{2}, \frac{N_f}{2}-f  \right].
\end{split}
\end{align} 
These intervals have the property that $\mc{I}_1 \cap \mc{I}_2$ and $\mc{I}_3 \cap \mc{I}_4$ can not both be non-zero. If $f>\frac{N_f}{2}$, the former is non-empty while the latter is empty and vice versa for $f\le\frac{N_f}{2}$. For regions where these mass intervals overlap, the preferred symmetry breaking scenario is the one which minimizes the effective potential. From eq. \eqref{eq:scenarios} it is easy to see that if $m<M$ then
\begin{equation*}
V_1 < V_2\text{   and   } V_4<V_3
\end{equation*}
while if $m>M$ we have the opposite. If $m=M$ the flavor group is no longer broken and we return to the phase diagram of \cite{Komargodski:2017keh}.

To proceed let us assume $m<M$. The regions in the $\tilde{m}-\tilde{M}$ plane for which symmetry breaking of either type will occur is (up to a factor of $4$)  are given by $\mc{I}_1\times \mathbb{R}^{m<M}$ and $\mathbb{R}^{m<M}\times\mc{I}_4$, respectively, where we defined 
\begin{equation*}
\mathbb{R}^{m<M}\equiv \{ (M,m) \in \mathbb{R}^2\, |\, m<M \}.
\end{equation*}
One can see that for $\tilde{M}/4 \in \mc{I}_4$, we have $\tilde{M}<0$ except for some small interval $[0,N_f/2-f]$ if $f<N_f/2$. Similarly, if $\tilde{m}/4 \in \mc{I}_1$, then $\tilde{m} >0$ except for some small interval $[N_f/2-f,0]$ if $f>N_f/2$. This shows that \textit{most} of the region where the $f$ fermions condense are for positive values of $\tilde{m}$ while the regions where the other $N_f-f$ fermions condense are for negative values of $\tilde{M}$! And since we restrict our attention to the regions $m<M$, we know that in the first quadrant, we get Grassmannians of the form $Gr(p,f)$ and in the third quadrant, we get Grassmannians of the form $Gr(q,N_f-f)$. In both cases, as you tune $\tilde{m}$ or $\tilde{M}$, you encounter a series of first order phase transitions which bring you through each Grassmannian until you wind up in the asymptotic, topological phase as in \cite{Komargodski:2017keh}. We can repeat the analysis for $M>m$ and find a similar phenomenon, except now we get Grassmannians of the form $Gr(p,f)$ for negative values of $\tilde{m}$ and Grassmannians of the form $Gr(q,N_f-f)$ for positive values of $\tilde{M}$. 

\begin{figure}[H] 
\centering
\subfloat[Phase diagram for $f<N_f/2$]{
\includegraphics[scale=.3 ]{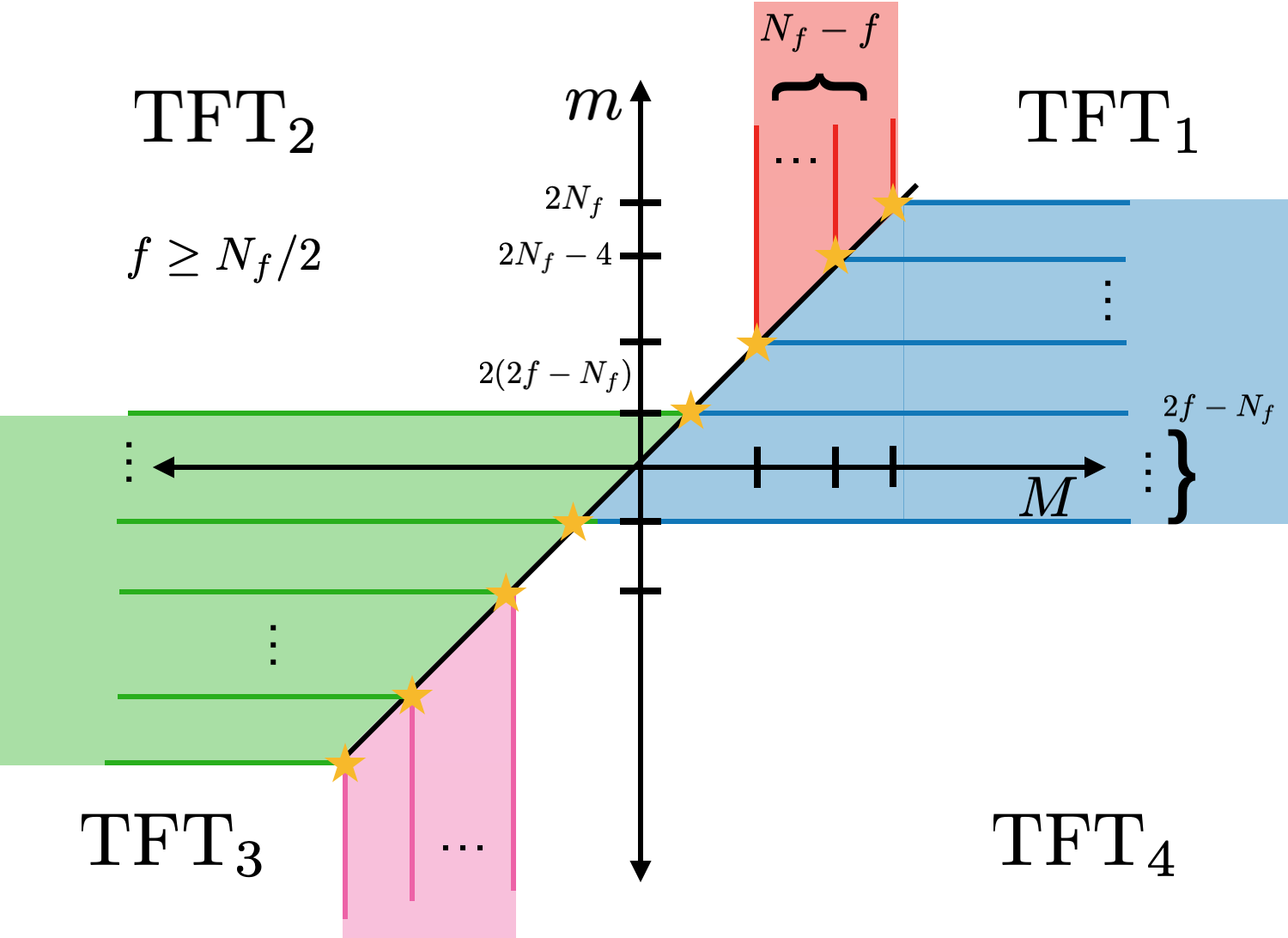}
}\subfloat[Phase diagram for $f\ge N_f/2$]{
\includegraphics[scale=.3 ]{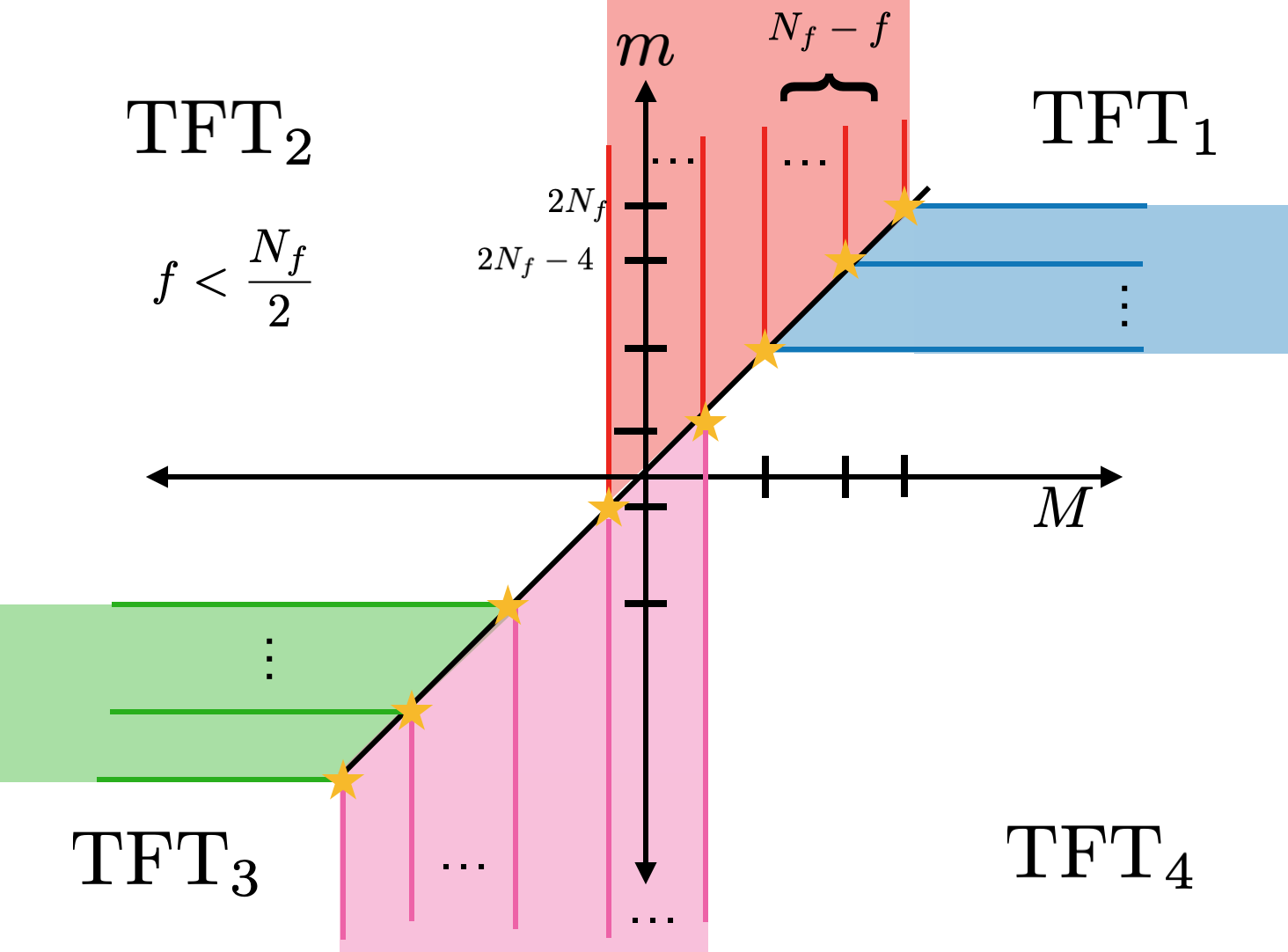}
}
\caption{Phase diagrams for QCD$_3$ with an explicitly broken flavor symmetry. The various TFTs shown are $SU(N)$ Chern-Simons gauge theory with level $ k+\frac{1}{2} (\text{sgn}(m)f+\text{sgn}(M)(N_f-f))$. The shaded regions represent the Grassmannians as explained in eq. \eqref{eq:explain}}\label{fig:diag}
\end{figure}

The resulting phase diagram is given in figure \ref{fig:diag} Each line represents a transition between Grassmannians, and they are colored according to the component of $\partial \mc{D}$ on which the transitions occur. Namely, blue corresponds to $\partial \mc{D}_1$, pink to $\partial \mc{D}_4$, green to $\partial \mc{D}_2$ and red to $\partial \mc{D}_3$. The low energy theory in each region is

\begin{align} \label{eq:explain}
\begin{split}
\text{Blue:}&\quad Gr(p,f) \otimes SU(N)_{\frac{N_f}{2}-p} \text{ with } p=\frac{N_f}{2} -\llbracket \frac{\tilde{m}}{4} \rrbracket\\
\text{Pink:}& \quad Gr(q,N_f-f) \otimes SU(N)_{N_f/2-f-q}\text{ with } q=\frac{N_f}{2}-f -\llbracket \frac{\tilde{M}}{4} \rrbracket \\
\text{Green:} &\quad Gr(p,f) \otimes SU(N)_{f-\frac{N_f}{2}-p}\text{ with } p=f-\frac{N_f}{2} -\llbracket \frac{\tilde{m}}{4} \rrbracket \\
\text{Red:} & \quad Gr(q,N_f-f) \otimes SU(N)_{N_f/2-q} \text{ with }  q=\frac{N_f}{2} -\llbracket \frac{\tilde{M}}{4} \rrbracket.
\end{split}
\end{align} 
These are subject to the isomorphism $Gr(M,N)\simeq Gr(N-M,N)$. Since we have chosen $p$ and $q$ to represent the number of $+1$ eigenvalues of our condensate, this isomorphism simply puts things in terms of the number of $-1$ eigenvalues. Physically these are equivalent. The black diagonal represents the flavor symmetric case studied in \cite{Komargodski:2017keh} where the stars represent the phase transitions along that line. In the flavor symmetric case, these were special points where two Grassmannian phases become degenerate. Now, they are phases where six different Grassmannians become degenerate. The vacuum one winds up in depends on the direction in the $M-m$ plane that you approach this point. For instance, if you tune along the $m<M$ from above or below, you wind up in a vacuum of the form $Gr(p,f)$, while if you tune along the $m=M$ from above or below you wind up in in $Gr(p,N_f)$. In general, if we have $n$ different masses there will be $2n$ degenerate vacua at these special values of the mass. 

One interesting part of the phase diagram in figure \ref{fig:diag} is the part where $\mc{I}_1 \cap \mc{I}_2 \ne \emptyset$ (or $\mc{I}_3 \cap \mc{I}_4\ne \emptyset$ if $f<N_f/2$). As we mentioned previously, these low energy theories are determined by which mass is larger. Interestingly, although perhaps not surprisingly, these theories are not equivalent and are related by a time-reversal-symmetry transformation. To see this note that the in the region of overlap, the blue Grassmannians are of the form $Gr(f-p,f)\otimes SU(N)_{N_f/2-f+p}$ while the green Grassmannians are of the form $Gr(p,f)\otimes SU(N)_{f-N_f/2-p}$. The Grassmannians are equivalent due to the isomorphism we discussed above, but the the Chern-Simons TFTs are related to each other by a time reversal operation! Thus, as one smoothly varies $m$ from $M-\epsilon$ to $m+\epsilon$, one should encounter a domain wall with extra light matter, even though the massless degrees of freedom on both sides are equivalent. Constructing such domain wall solutions is interesting but is beyond the scope of this work. If one includes a non-zero Chern-Simons level, then the TFTs are no longer related by a time reversal symmetry, but one should get a domain wall solution regardless.

In addition to this analytic derivation, we also include some numerical evidence for a specific choice of $N_f$ and $f$. We find perfect agreement with our analysis in this particular case, as well as for numerous other choices. For brevity we only include the one. 

\subsubsection{Matching Onto the Finite N Solution}
As a check, we would like to match onto the flavor broken phase diagram of \cite{Baumgartner:2019frr, Argurio:2019tvw}. This is not entirely possible, however, due to the fact that the adding a Chern-Simons term does not alter the shape of the phase diagram, only the location in the plane where the transitions occurs. However, as noted in \cite{Armoni:2019lgb} the finite $N$ symmetry broken phase is still present in the large $N$ solution and is distinguished by the fact that it is the phase which has vanishing Chern-Simons level for it's decoupled TFT. When $k=0$, this occurs at $p=N_f/2$ giving the symmetry broken phase first discussed in \cite{Vafa:1984xh,Vafa:1983tf}. At non-zero $k$, this occurs when $p=k+N_f/2$, giving the phase discussed in \cite{Komargodski:2017keh}. Thus, one consistency check we can perform is to see if the classical flavor-broken Grassmannians discussed in \cite{Baumgartner:2019frr, Argurio:2019tvw} (and shown in \ref{uneqDiags}) are accompanied by  $SU(N)_0$ for the relevant parameter regimes. 

To proceed, we find for what values of $p$ or $q$ the Chern-Simons level vanish, and find the values of $k$ for which this confining theory exists. We get: 
\begin{align} \label{eq:explain}
\begin{split}
\text{Blue: }\,\,& p=k+\frac{N_f}{2}\,\, \text{ exists when }\,\,-\frac{N_f}{2}\le k \le f-\frac{N_f}{2} \\ 
\text{Pink:}\,\, & q=k+\frac{N_f}{2}-f \,\,\text{ exists when } \,\, f-\frac{N_f}{2}\le k \le \frac{N_f}{2} \\
\text{Green:} \,\, & p=k+f-\frac{N_f}{2}\,\,\text{ exists when } \,\, \frac{N_f}{2}-f\le k \le \frac{N_f}{2} \\
\text{Red:} \,\, & q=k+\frac{N_f}{2}-f\,\,\text{ exists when } \,\, -\frac{N_f}{2}\le k \le \frac{N_f}{2}-f.
\end{split}
\end{align} 
Comparing these ranges to those discussed in section \ref{unequal}, we find the following: 

\begin{enumerate}[label=\roman*] 
\centering
 \item[i.)] $f-\frac{N_f}{2}<k$, $\frac{N_f}{2}-f<k$--- Green+ Pink
 \item[ii.)] $f-\frac{N_f}{2}<k$, $\frac{N_f}{2}-f=k$---Pink only
 \item[iii.)] $f-\frac{N_f}{2}<k$, $\frac{N_f}{2}-f>k$---Pink+Red
 \item[iv.)] $f-\frac{N_f}{2}=k$, $\frac{N_f}{2}-f>k$---Red Only
\item[v.)] $f-\frac{N_f}{2}>k$, $\frac{N_f}{2}-f>k$---Blue + Red
\item[vi.)] $f-\frac{N_f}{2}=k$,$\frac{N_f}{2}-f=k$---None
\end{enumerate}
As you can see, this perfectly matches with the phase diagrams in \cite{Baumgartner:2019frr, Argurio:2019tvw} and figure \ref{uneqDiags}. 

\section{Scalar Potentials and Double Condensation} \label{scalars}
When we pass to the scalar side of the large N $QCD$ duality we encounter a number of difficulties. Specifically, the authors of \cite{Armoni:2019lgb} have argued that the usual quartic potential at finite N will not accurately capture the phase structure of large N QCD$_3$. They go on to construct a sextic potential at $\mc{O}(N)$ and $\mc{O}(1)$ which can reproduce the desired effects: 
\begin{align}
V_{\mc{O}(N)} \sim & \sum_i y_i^2(y_i^2-1)^2 \label{eq:scalOrdN}\\ 
V_{\mc{O}(1)} \sim & \sum_{i,j} (2y_i^2-1)(2y_j^2-1) \label{eq:scalOrd1}
\end{align}
where $y_i$ is the eigenvalue for the scalar vacuum expectation value. However, as was argued in \cite{Armoni:2019lgb}, the coupling constants associated with eqs. \eqref{eq:scalOrdN} and \eqref{eq:scalOrd1} must be fine tuned to ensure consistency with the fermionic side. This is an unfortunate artifact, but the fact that it can be done for some choice of coefficients is reassuring. 

When we break the flavor symmetry to the $U(f)\times U(N_f-f)$ subgroup we generally expect to operators which respect this new symmetry to be generated along the RG flow. On the fermion side of the duality, it is not clear \textit{a priori} what these would be, since quartic potentials are classically irrelevant in 2+1 d. For this we turn to the scalar side to see these effects. For simplicity we stick to $U(f)\times U(N_f-f)$ quartic potentials. 

Let $\phi_1$ represent the first $f$ scalars and $\phi_2$ the remaining $N_f-f$. Then, the two quartic $U(f)\times U(N_f-f)$ invariant potentials are 
\begin{align}
\mathcal{L} \supset &\,\, \lambda\phi_{1I}^{\dagger b}\phi_{1I}^{a}\phi_{2J}^{\dagger a}\phi_{2J}^{b} \label{eq:scalPotBreak1}\\ 
\supset &\,\,  \gamma\phi_{1I}^{\dagger a}\phi_{1I}^{a}\phi_{2J}^{\dagger b}\phi_{2J}^{b}. \label{eq:scalPotBreak2}
\end{align} 
Both of these involve double traces over the flavor indicies, while eq. \eqref{eq:scalPotBreak1} has a single trace over the gauge indicies. The index structure of eq. \eqref{eq:scalPotBreak2} indicates it should be part of the $\mc{O}(1)$ potential, since it involves separate gauge and flavor traces. Diagrams with this interaction will be included in the annulus diagram as discussed in section \ref{sec:2}.

\begin{figure}
\centering
\includegraphics[scale=.5]{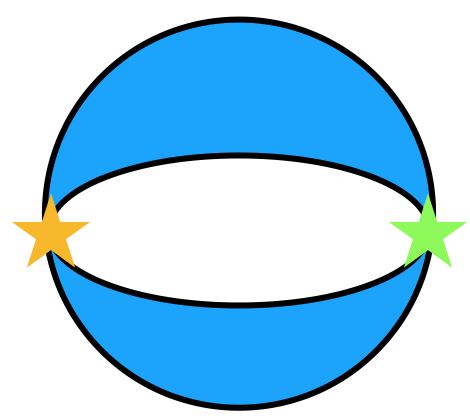}
\caption{Diagram associated to the $U(f)\times U(N_f-f)$ invariant term discussed in section \ref{scalars}}\label{fig:eye}
\end{figure}

The more interesting term is eq. \eqref{eq:scalPotBreak1}. Terms of this type have been discussed before in \cite{Aitken:2018cvh,Baumgartner:2019frr}. Heuristically this term behaves as follows: when one set of scalars gets a vev from a negative mass deformation this term induces a mass for some of the components of the other group of scalars. In \cite{Baumgartner:2019frr} it was argued that if this term is large enough it could prevent double condensation, since any negative mass deformation could not cancel the off the contribution from the induced mass. The gauge and flavor index structure makes it difficult to determine at which order this type of term should enter since diagrams containing this interaction are neither disk or annulus diagrams but instead are ``eyeball"  type diagrams as in figure \ref{fig:eye}. Since these are, in a sense, `` in between" a disk and annulus, we examine the effect of including this term at order $N$ and order 1. 

Consider the following forms of the vevs for the scalars. If the rank of the gauge group is larger than the (broken) flavor group, the vev will look like: 
\begin{equation}\label{eq:vevGen}
\phi_{1}^{aI}=\left(\begin{array}{c}
diag(x_{1},...,x_{f})\\
0
\end{array}\right),\;\phi_{2}^{aI}=\left(\begin{array}{c}
diag(y_{1},...,y_{N_{f}-f})\\
0
\end{array}\right).
\end{equation}
When this is multiplied out according to \eqref{eq:scalPotBreak1}, we wind up with a term in the effective potential which looks like 
\begin{equation}
V\supset \,\, \lambda \sum_{i=1}^F x_i^2y_i^2 
\end{equation}
where $F\equiv \text{min}(f,N_f-f)$. If we include this term in the $O(1)$ potential, then we would be looking for the optimal configuration of 0's and 1's along the diagonal. This would add a term like $\lambda \text{min}(p,q)$, where now $p$ and $q$ refer to the number of 1's on the diagonals of \eqref{eq:vevGen}.  to the minimization problem we have discussed. Again, there is no minimum in the interior of our domain in the $p-q$ plane and the absolute minimum must live on the boundary. On the components of the boundary where either $p$ or $q$ are zero, this term disappears and the analysis is the same as it was without the term. If we are on the other two boundaries, this term becomes something like $\lambda p$ or $\lambda q$ and will serve to simply shift the mass by $\tilde{m}\to \tilde{m}+\lambda/2$. Evidently, in either case the analysis is identical to the inclusion of a Chern-Simons term. 

The more interesting case is when we include this term in the $\mc{O}(N)$ potential. In which case, we must minimize
\begin{equation}\label{eq:newTerm}
V_{\mc{O}(N)} \sim \sum_{i=1}^{f}y_{i}^{2}(y_{i}^{2}-1)^{2}+\sum_{i=1}^{N_{f}-f}x_{i}^{2}(x_{i}^{2}-1)^{2}+\lambda\sum_{i=1}^{min(f,N_{f}-f)}y_{i}^{2}x_{i}^{2}.
\end{equation}
Again there are minimum at $x_,y_i=0,1$ however the eigenvalues are now correlated.  For generic $\lambda$ the minimum of eq. \eqref{eq:newTerm} occur at $(x_i,y_i)=(0,1),(1,0)$ and $(0,0)$. Thus, if we choose the first $p$ of the $x$'s to be $1$, we must take the first $p$ of the $y$'s to be 0. However, if we choose the first $p$ $x$'s to be 0, then we can choose the corresponding $y$'s to be $0$ or $1$. Without adding any mass deformations, we thus have many degenerate Grassmannians which, generically, are of the form 
\begin{equation}
Gr(p,f)\times Gr(q,N_f-f)
\end{equation}
where $N_f-f-p\le q \le N_f-f$. If $q$ does not fall within this range, we wind up with single condensation. Also, if we give $\phi_1$ a positive mass deformation, it forces all the $y$'s to 0 and we wind up with degenerate vacua as in section \ref{eq:meat}. If we give them a negative mass deformation, we force all the $y$'s to be 1. This then restricts the allowed range of $q$ and we do not have as many degenerate vacua. 

One could further ask what happens if we include eq. \eqref{eq:scalPotBreak2} at $\mc{O}(1)$ in addition to eq. \eqref{eq:scalPotBreak1} at $\mc{O}(N)$. This gets messy and so we will not perform a full analysis here, but we do note that for generic values of $\lambda$ the minimum again occurs on the boundary of the $p-q$ plane. However, there do exist regions of the $m-M$ phase diagram where double condensation at $\mc{O}(1)$ can occur for finely tuned values of $\gamma$. This is true with or without the inclusion of the $\lambda$ term---the $\lambda$ term only serves to restrict the number of Grassmannians one is allowed to traverse due to the correlation of the scalar vev eigenvalues. So again, we could engineer phases on the scalar side of the theory where double condensation can occur in the phase diagram. It is difficult to know if this symmetry breaking actually occurs for QCD$_3$ unless we can accurately calculate the anomalous dimensions of $\gamma$ and $\lambda$. For now, we only note that with the inclusion of these symmetry breaking potentials, it is possible for both flavors of fermions to condense simultaneously.

\section{Conclusion}\label{sec:con}
In this work we studied the large $N$ limit of QCD$_3$ as a function of two distinct masses. We find that there exists multiple Grassmannians separated by a series of first order phase transitions in certain portions of the phase diagram. These are consistent with the phase diagrams at finite $N$ laid out in \cite{Baumgartner:2019frr, Argurio:2019tvw}. We find a plethora of interesting physics, including multiple degenerate Grassmannians at special values on the mass, degenerate symmetric and symmetry broken phases, and lack of double condensation for unmodified scalar potentials. Additionally, we consider the effect of quartic interactions which may possibly be generated along the RG flow and find that double condensation is possible for finely tuned values of the interaction strength. 

An interesting extension to this work is to find a more rigorous argument for the lack of double condensation without the modified scalar potentials. Perhaps there is a deeper reason for this other than the lack of solution for $p$ and $q$ on the interior of the domain $\mathcal{D}$. It does not seem to be forbidden by an f-theorem or anomaly matching type argument, at least not that the author can work out. The generation of $U(f)\times U(N_f-f)$ terms in the scalar potential may be indication that no such general argument exists. Still, it might be worth investigating further. Additionally, it would be interesting to attempt to compute the anomalous dimensions of these extra terms in the scalar potential to determine if they even get generated along the RG flow. If not, this could be further indication of a more fundamental reason for lack of double condensation. 

The rich structure of this phase diagram lends itself nicely to the study of domain walls between Grassmannians. For instance, fix the magnitude of the mass of $f-1$ of the flavors and tuning the mass of the remaining fermion from $m\to M$, we encounter a domain wall with extra light matter living on it. What is the nature of this matter? Does it couple to the Grassmannians? These are questions which are beyond the scope of this work but seem interesting enough to warrant further investigation.

\appendix
\section*{Acknowledgments}
We would like to thank Andreas Karch for useful advice and comments on this manuscript. We also thank Tyler Ellison an Ryan Lanzetta enlightening discussions. This work was supported, in part, by the U.S.~Department of Energy under Grant No.~DE-SC0011637.

\section{Numerical Evidence for the Phase Diagram}\label{sec:append}
\begin{figure}[h]
\centering
\includegraphics[scale=.5]{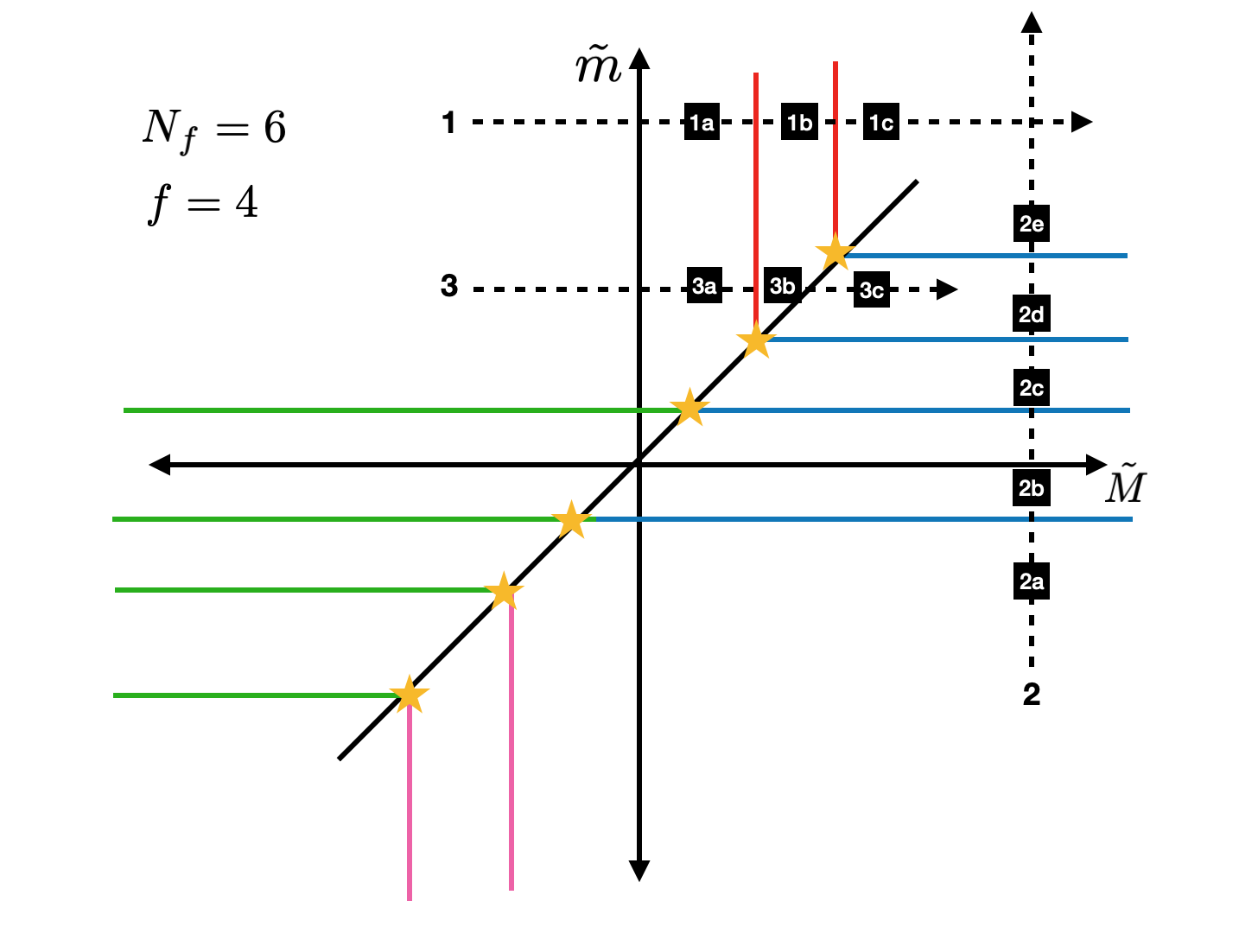}
\caption{Phase diagram for $N_f=6,f=4$ with $k=0$. Various effective potential at the indicated points along trajectories 1,2 and 3 shown are given in tables \ref{NumFixedM}, \ref{NumFixm} and \ref{NumsMids}.}\label{NumFig}
\end{figure}

Here we present a subset of numerical results which corroborate the phase diagram discussed in the main text.  For simplicity, we choose $N_f=6$ and $f=4$ and $k=0$. We stress that these choices do not impact the qualitative results, only the location and number of phase transitions, as well as the level of the accompanying Chern-Simons TFT. To demonstrate the lack of double condensation, we fix one mass and tune the other. This brings us along the three trajectories shown in figure \ref{NumFig}. Note that as we tune the mass, the minimum value of the effective potential remains along the boundary of the table. This corresponds to values of $p$ or $q$ which exist along the boundary of the domain $\mc{D}$. The trajectories along fixed, positive $\tilde{m}\equiv  \frac{mN}{4\Lambda \Delta} $ and $\tilde{M}\equiv  \frac{MN}{4\Lambda \Delta} $ are related to those trajectories along fixed negative $\tilde{m}$ and $\tilde{M}$ by a time reversal transformation. This maps $p\to f-p$ and $q\to N_f-f-q$, which serves to invert the rows and reverse the columns of tables \ref{NumFixedM},\ref{NumFixm} and \ref{NumsMids}

\begin{table}[!ht]
\centering
\subfloat[$\tilde{M}=3$ with vacuum $SU(N)_{1}$]{
\begin{tabular}{llll}
                                                                          & \cellcolor[HTML]{C0C0C0}$q=0$ & \cellcolor[HTML]{C0C0C0}1 & \cellcolor[HTML]{C0C0C0}2                        \\ \cline{2-4} 
\multicolumn{1}{l|}{\cellcolor[HTML]{C0C0C0}{\color[HTML]{333333} \,$p=0$}} & \multicolumn{1}{l|}{-18}      & \multicolumn{1}{l|}{-32}  & \multicolumn{1}{l|}{\cellcolor[HTML]{FE0000}-38} \\ \cline{2-4} 
\multicolumn{1}{l|}{\cellcolor[HTML]{C0C0C0}{\color[HTML]{333333} \qquad 1}}     & \multicolumn{1}{l|}{-14}      & \multicolumn{1}{l|}{-20}  & \multicolumn{1}{l|}{-18}                         \\ \cline{2-4} 
\multicolumn{1}{l|}{\cellcolor[HTML]{C0C0C0}{\color[HTML]{333333} \qquad 2}}     & \multicolumn{1}{l|}{-2}       & \multicolumn{1}{l|}{0}    & \multicolumn{1}{l|}{10}                          \\ \cline{2-4} 
\multicolumn{1}{l|}{\cellcolor[HTML]{C0C0C0}{\color[HTML]{333333} \qquad3}}     & \multicolumn{1}{l|}{18}       & \multicolumn{1}{l|}{28}   & \multicolumn{1}{l|}{46}                          \\ \cline{2-4} 
\multicolumn{1}{l|}{\cellcolor[HTML]{C0C0C0}{\color[HTML]{333333} \qquad 4}}     & \multicolumn{1}{l|}{46}       & \multicolumn{1}{l|}{64}   & \multicolumn{1}{l|}{90}                          \\ \hline
\end{tabular}
}\hfill
\subfloat[$\tilde{M}=7$ with vacuum $Gr(1,2)\otimes SU(N)_2$]{
\begin{tabular}{llll}
                                                                          & \cellcolor[HTML]{C0C0C0}$q=0$ & \cellcolor[HTML]{C0C0C0}1                        & \cellcolor[HTML]{C0C0C0}2                        \\ \cline{2-4} 
\multicolumn{1}{l|}{\cellcolor[HTML]{C0C0C0}{\color[HTML]{333333} \,$p=0$}} & \multicolumn{1}{l|}{-26}      & \multicolumn{1}{l|}{\cellcolor[HTML]{FE0000}-32} & \multicolumn{1}{l|}{\cellcolor[HTML]{FFFFFF}-30} \\ \cline{2-4} 
\multicolumn{1}{l|}{\cellcolor[HTML]{C0C0C0}{\color[HTML]{333333} \qquad 1}}     & \multicolumn{1}{l|}{-22}      & \multicolumn{1}{l|}{-20}                         & \multicolumn{1}{l|}{-10}                         \\ \cline{2-4} 
\multicolumn{1}{l|}{\cellcolor[HTML]{C0C0C0}{\color[HTML]{333333} \qquad 2}}     & \multicolumn{1}{l|}{-10}      & \multicolumn{1}{l|}{0}                           & \multicolumn{1}{l|}{18}                          \\ \cline{2-4} 
\multicolumn{1}{l|}{\cellcolor[HTML]{C0C0C0}{\color[HTML]{333333} \qquad 3}}     & \multicolumn{1}{l|}{10}       & \multicolumn{1}{l|}{28}                          & \multicolumn{1}{l|}{54}                          \\ \cline{2-4} 
\multicolumn{1}{l|}{\cellcolor[HTML]{C0C0C0}{\color[HTML]{333333} \qquad 4}}     & \multicolumn{1}{l|}{38}       & \multicolumn{1}{l|}{64}                          & \multicolumn{1}{l|}{98}                          \\ \hline
\end{tabular}
} \hfill
\subfloat[$\tilde{M}=11$ with vacuum $SU(N)_3$]{
\begin{tabular}{llll}
                                                                          & \cellcolor[HTML]{C0C0C0}$q=0$                    & \cellcolor[HTML]{C0C0C0}1                        & \cellcolor[HTML]{C0C0C0}2                        \\ \cline{2-4} 
\rowcolor[HTML]{FFFFFF} 
\multicolumn{1}{l|}{\cellcolor[HTML]{C0C0C0}{\color[HTML]{333333} \,$p=0$}} & \multicolumn{1}{l|}{\cellcolor[HTML]{FE0000}-34} & \multicolumn{1}{l|}{\cellcolor[HTML]{FFFFFF}-32} & \multicolumn{1}{l|}{\cellcolor[HTML]{FFFFFF}-22} \\ \cline{2-4} 
\multicolumn{1}{l|}{\cellcolor[HTML]{C0C0C0}{\color[HTML]{333333} \qquad 1}}     & \multicolumn{1}{l|}{-30}                         & \multicolumn{1}{l|}{-20}                         & \multicolumn{1}{l|}{-2}                          \\ \cline{2-4} 
\multicolumn{1}{l|}{\cellcolor[HTML]{C0C0C0}{\color[HTML]{333333} \qquad 2}}     & \multicolumn{1}{l|}{-18}                         & \multicolumn{1}{l|}{0}                           & \multicolumn{1}{l|}{26}                          \\ \cline{2-4} 
\multicolumn{1}{l|}{\cellcolor[HTML]{C0C0C0}{\color[HTML]{333333} \qquad 3}}     & \multicolumn{1}{l|}{2}                           & \multicolumn{1}{l|}{28}                          & \multicolumn{1}{l|}{62}                          \\ \cline{2-4} 
\multicolumn{1}{l|}{\cellcolor[HTML]{C0C0C0}{\color[HTML]{333333} \qquad 4}}     & \multicolumn{1}{l|}{30}                          & \multicolumn{1}{l|}{64}                          & \multicolumn{1}{l|}{106}                         \\ \hline
\end{tabular}
}
\caption{Various values of the effective potential with $\tilde{m}=12$. This follows trajectory 1 as shown in \ref{NumFig}. The minimum is shown in red.}
 \label{NumFixedM}
\end{table}

\begin{table}[h]
\centering
\subfloat[$\tilde{m}=-5$ with vacuum $SU(N)_{-1}$]{
\begin{tabular}{llll}
                                                                          & \cellcolor[HTML]{C0C0C0}$q=0$                    & \cellcolor[HTML]{C0C0C0}1                       & \cellcolor[HTML]{C0C0C0}2                       \\ \cline{2-4} 
\rowcolor[HTML]{FFFFFF} 
\multicolumn{1}{l|}{\cellcolor[HTML]{C0C0C0}{\color[HTML]{333333}  \,$p=0$}} & \multicolumn{1}{l|}{\cellcolor[HTML]{FFFFFF}26}  & \multicolumn{1}{l|}{\cellcolor[HTML]{FFFFFF}36} & \multicolumn{1}{l|}{\cellcolor[HTML]{FFFFFF}54} \\ \cline{2-4} 
\multicolumn{1}{l|}{\cellcolor[HTML]{C0C0C0}{\color[HTML]{333333} \qquad 1}}     & \multicolumn{1}{l|}{-4}                          & \multicolumn{1}{l|}{14}                         & \multicolumn{1}{l|}{40}                         \\ \cline{2-4} 
\multicolumn{1}{l|}{\cellcolor[HTML]{C0C0C0}{\color[HTML]{333333} \qquad 2}}     & \multicolumn{1}{l|}{-26}                         & \multicolumn{1}{l|}{0}                          & \multicolumn{1}{l|}{34}                         \\ \cline{2-4} 
\multicolumn{1}{l|}{\cellcolor[HTML]{C0C0C0}{\color[HTML]{333333}\qquad 3}}     & \multicolumn{1}{l|}{-40}                         & \multicolumn{1}{l|}{-6}                         & \multicolumn{1}{l|}{36}                         \\ \cline{2-4} 
\multicolumn{1}{l|}{\cellcolor[HTML]{C0C0C0}{\color[HTML]{333333}\qquad 4}}     & \multicolumn{1}{l|}{\cellcolor[HTML]{FE0000}-46} & \multicolumn{1}{l|}{-4}                         & \multicolumn{1}{l|}{46}                         \\ \hline
\end{tabular}
}\hfill
\subfloat[$\tilde{m}=-1$ with vacuum $Gr(3,4)\otimes SU(N)_0$]{\begin{tabular}{llll}
                                                                          & \cellcolor[HTML]{C0C0C0}$q=0$                                           & \cellcolor[HTML]{C0C0C0}1                       & \cellcolor[HTML]{C0C0C0}2                       \\ \cline{2-4} 
\rowcolor[HTML]{FFFFFF} 
\multicolumn{1}{l|}{\cellcolor[HTML]{C0C0C0}{\color[HTML]{333333} \,$p=0$}} & \multicolumn{1}{l|}{\cellcolor[HTML]{FFFFFF}10}                         & \multicolumn{1}{l|}{\cellcolor[HTML]{FFFFFF}20} & \multicolumn{1}{l|}{\cellcolor[HTML]{FFFFFF}38} \\ \cline{2-4} 
\multicolumn{1}{l|}{\cellcolor[HTML]{C0C0C0}{\color[HTML]{333333} \qquad 1}}     & \multicolumn{1}{l|}{-12}                                                & \multicolumn{1}{l|}{6}                          & \multicolumn{1}{l|}{32}                         \\ \cline{2-4} 
\multicolumn{1}{l|}{\cellcolor[HTML]{C0C0C0}{\color[HTML]{333333}\qquad 2}}     & \multicolumn{1}{l|}{-26}                                                & \multicolumn{1}{l|}{0}                          & \multicolumn{1}{l|}{34}                         \\ \cline{2-4} 
\multicolumn{1}{l|}{\cellcolor[HTML]{C0C0C0}{\color[HTML]{333333}\qquad 3}}     & \multicolumn{1}{l|}{\cellcolor[HTML]{FE0000}{\color[HTML]{333333} -32}} & \multicolumn{1}{l|}{2}                          & \multicolumn{1}{l|}{44}                         \\ \cline{2-4} 
\multicolumn{1}{l|}{\cellcolor[HTML]{C0C0C0}{\color[HTML]{333333}\qquad 4}}     & \multicolumn{1}{l|}{\cellcolor[HTML]{FFFFFF}-30}                        & \multicolumn{1}{l|}{12}                         & \multicolumn{1}{l|}{62}                         \\ \hline
\end{tabular}
}\hfill
\subfloat[$\tilde{m}=3$ with vacuum $Gr(2,4)\otimes SU(N)_1$]{
\begin{tabular}{llll}
                                                                          & \cellcolor[HTML]{C0C0C0}$q=0$                                           & \cellcolor[HTML]{C0C0C0}1                      & \cellcolor[HTML]{C0C0C0}2                       \\ \cline{2-4} 
\rowcolor[HTML]{FFFFFF} 
\multicolumn{1}{l|}{\cellcolor[HTML]{C0C0C0}{\color[HTML]{333333} $p=0$}} & \multicolumn{1}{l|}{\cellcolor[HTML]{FFFFFF}-6}                         & \multicolumn{1}{l|}{\cellcolor[HTML]{FFFFFF}4} & \multicolumn{1}{l|}{\cellcolor[HTML]{FFFFFF}22} \\ \cline{2-4} 
\multicolumn{1}{l|}{\cellcolor[HTML]{C0C0C0}{\color[HTML]{333333} 1}}     & \multicolumn{1}{l|}{-20}                                                & \multicolumn{1}{l|}{-2}                        & \multicolumn{1}{l|}{24}                         \\ \cline{2-4} 
\multicolumn{1}{l|}{\cellcolor[HTML]{C0C0C0}{\color[HTML]{333333} 2}}     & \multicolumn{1}{l|}{\cellcolor[HTML]{FE0000}{\color[HTML]{333333} -26}} & \multicolumn{1}{l|}{0}                         & \multicolumn{1}{l|}{34}                         \\ \cline{2-4} 
\multicolumn{1}{l|}{\cellcolor[HTML]{C0C0C0}{\color[HTML]{333333} 3}}     & \multicolumn{1}{l|}{ -24} & \multicolumn{1}{l|}{10}                        & \multicolumn{1}{l|}{52}                         \\ \cline{2-4} 
\multicolumn{1}{l|}{\cellcolor[HTML]{C0C0C0}{\color[HTML]{333333} 4}}     & \multicolumn{1}{l|}{-14}                        & \multicolumn{1}{l|}{28}                        & \multicolumn{1}{l|}{78}                         \\ \hline
\end{tabular}
}

\subfloat[$\tilde{m}=7$ with vacuum $Gr(1,4)\otimes SU(N)_2$]{
\begin{tabular}{l
>{\columncolor[HTML]{FFFFFF}}l ll}
                                                                          & \cellcolor[HTML]{C0C0C0}$q=0$                                           & \cellcolor[HTML]{C0C0C0}1                        & \cellcolor[HTML]{C0C0C0}2                      \\ \cline{2-4} 
\multicolumn{1}{l|}{\cellcolor[HTML]{C0C0C0}{\color[HTML]{333333} \,$p=0$}} & \multicolumn{1}{l|}{\cellcolor[HTML]{FFFFFF}-22}                        & \multicolumn{1}{l|}{\cellcolor[HTML]{FFFFFF}-12} & \multicolumn{1}{l|}{\cellcolor[HTML]{FFFFFF}6} \\ \cline{2-4} 
\multicolumn{1}{l|}{\cellcolor[HTML]{C0C0C0}{\color[HTML]{333333}\qquad 1}}     & \multicolumn{1}{l|}{\cellcolor[HTML]{FE0000}-28}                        & \multicolumn{1}{l|}{-10}                         & \multicolumn{1}{l|}{16}                        \\ \cline{2-4} 
\multicolumn{1}{l|}{\cellcolor[HTML]{C0C0C0}{\color[HTML]{333333}\qquad 2}}     & \multicolumn{1}{l|}{ -26} & \multicolumn{1}{l|}{0}                           & \multicolumn{1}{l|}{34}                        \\ \cline{2-4} 
\multicolumn{1}{l|}{\cellcolor[HTML]{C0C0C0}{\color[HTML]{333333} \qquad 3}}     & \multicolumn{1}{l|}{-16} & \multicolumn{1}{l|}{18}                          & \multicolumn{1}{l|}{60}                        \\ \cline{2-4} 
\multicolumn{1}{l|}{\cellcolor[HTML]{C0C0C0}{\color[HTML]{333333} \qquad 4}}     & \multicolumn{1}{l|}{2}                          & \multicolumn{1}{l|}{44}                          & \multicolumn{1}{l|}{94}                        \\ \hline
\end{tabular}
}\hspace{.1\textwidth}
\subfloat[$\tilde{m}=11$  with vacuum $SU(N)_3$]{
\begin{tabular}{llll}
                                                                          & \cellcolor[HTML]{C0C0C0}$q=0$                                           & \cellcolor[HTML]{C0C0C0}1                        & \cellcolor[HTML]{C0C0C0}2                        \\ \cline{2-4} 
\rowcolor[HTML]{FFFFFF} 
\multicolumn{1}{l|}{\cellcolor[HTML]{C0C0C0}{\color[HTML]{333333} \,$p=0$}} & \multicolumn{1}{l|}{\cellcolor[HTML]{FE0000}-38}                        & \multicolumn{1}{l|}{\cellcolor[HTML]{FFFFFF}-28} & \multicolumn{1}{l|}{\cellcolor[HTML]{FFFFFF}-10} \\ \cline{2-4} 
\multicolumn{1}{l|}{\cellcolor[HTML]{C0C0C0}{\color[HTML]{333333} \qquad 1}}     & \multicolumn{1}{l|}{-36}                                                & \multicolumn{1}{l|}{-18}                         & \multicolumn{1}{l|}{8}                           \\ \cline{2-4} 
\multicolumn{1}{l|}{\cellcolor[HTML]{C0C0C0}{\color[HTML]{333333} \qquad 2}}     & \multicolumn{1}{l|}{ -26} & \multicolumn{1}{l|}{0}                           & \multicolumn{1}{l|}{34}                          \\ \cline{2-4} 
\multicolumn{1}{l|}{\cellcolor[HTML]{C0C0C0}{\color[HTML]{333333}\qquad 3}}     & \multicolumn{1}{l|}{-8}  & \multicolumn{1}{l|}{26}                          & \multicolumn{1}{l|}{68}                          \\ \cline{2-4} 
\multicolumn{1}{l|}{\cellcolor[HTML]{C0C0C0}{\color[HTML]{333333}\qquad 4}}     & \multicolumn{1}{l|}{18}                         & \multicolumn{1}{l|}{60}                          & \multicolumn{1}{l|}{110}                         \\ \hline
\end{tabular}
}
\caption{Various values of the effective potential with $\tilde{M}=15$. This follows trajectory 2 as shown in \ref{NumFig}. The minimum is shown in red.}
\label{NumFixm}
\end{table}

\begin{table}
\centering
\subfloat[$\tilde{M}=3$ with vacuum $SU(N)_1$]{
\begin{tabular}{llll}
                                                                          & \cellcolor[HTML]{C0C0C0}$q=0$ & \cellcolor[HTML]{C0C0C0}1                        & \cellcolor[HTML]{C0C0C0}2                        \\ \cline{2-4} 
\multicolumn{1}{l|}{\cellcolor[HTML]{C0C0C0}{\color[HTML]{333333} \,$p=0$}} & \multicolumn{1}{l|}{-2}       & \multicolumn{1}{l|}{\cellcolor[HTML]{FFFFFF}-16} & \multicolumn{1}{l|}{\cellcolor[HTML]{FE0000}-22} \\ \cline{2-4} 
\multicolumn{1}{l|}{\cellcolor[HTML]{C0C0C0}{\color[HTML]{333333}\qquad 1}}     & \multicolumn{1}{l|}{-6}       & \multicolumn{1}{l|}{-12}                         & \multicolumn{1}{l|}{-10}                         \\ \cline{2-4} 
\multicolumn{1}{l|}{\cellcolor[HTML]{C0C0C0}{\color[HTML]{333333}\qquad 2}}     & \multicolumn{1}{l|}{-2}       & \multicolumn{1}{l|}{0}                           & \multicolumn{1}{l|}{10}                          \\ \cline{2-4} 
\multicolumn{1}{l|}{\cellcolor[HTML]{C0C0C0}{\color[HTML]{333333}\qquad 3}}     & \multicolumn{1}{l|}{10}       & \multicolumn{1}{l|}{20}                          & \multicolumn{1}{l|}{38}                          \\ \cline{2-4} 
\multicolumn{1}{l|}{\cellcolor[HTML]{C0C0C0}{\color[HTML]{333333}\qquad 4}}     & \multicolumn{1}{l|}{30}       & \multicolumn{1}{l|}{48}                          & \multicolumn{1}{l|}{74}                          \\ \hline
\end{tabular}
}\hfill
\subfloat[$\tilde{M}=7$ with vacuum $Gr(1,2) \otimes SU(N)_2$]{
\begin{tabular}{llll}
                                                                          & \cellcolor[HTML]{C0C0C0}$q=0$ & \cellcolor[HTML]{C0C0C0}1                        & \cellcolor[HTML]{C0C0C0}2                        \\ \cline{2-4} 
\multicolumn{1}{l|}{\cellcolor[HTML]{C0C0C0}{\color[HTML]{333333} \,$p=0$}} & \multicolumn{1}{l|}{-10}      & \multicolumn{1}{l|}{\cellcolor[HTML]{FE0000}-16} & \multicolumn{1}{l|}{\cellcolor[HTML]{FFFFFF}-14} \\ \cline{2-4} 
\multicolumn{1}{l|}{\cellcolor[HTML]{C0C0C0}{\color[HTML]{333333}\qquad 1}}     & \multicolumn{1}{l|}{-14}      & \multicolumn{1}{l|}{-12}                         & \multicolumn{1}{l|}{-2}                          \\ \cline{2-4} 
\multicolumn{1}{l|}{\cellcolor[HTML]{C0C0C0}{\color[HTML]{333333}\qquad 2}}     & \multicolumn{1}{l|}{-10}      & \multicolumn{1}{l|}{0}                           & \multicolumn{1}{l|}{18}                          \\ \cline{2-4} 
\multicolumn{1}{l|}{\cellcolor[HTML]{C0C0C0}{\color[HTML]{333333}\qquad 3}}     & \multicolumn{1}{l|}{2}        & \multicolumn{1}{l|}{20}                          & \multicolumn{1}{l|}{46}                          \\ \cline{2-4} 
\multicolumn{1}{l|}{\cellcolor[HTML]{C0C0C0}{\color[HTML]{333333}\qquad 4}}     & \multicolumn{1}{l|}{22}       & \multicolumn{1}{l|}{48}                          & \multicolumn{1}{l|}{82}                          \\ \hline
\end{tabular}
}\hfill
\subfloat[$\tilde{M}=11$ with vacuum $Gr(1,4)\otimes SU(N)_2$]{
\begin{tabular}{llll}
                                                                          & \cellcolor[HTML]{C0C0C0}$q=0$                                           & \cellcolor[HTML]{C0C0C0}1 & \cellcolor[HTML]{C0C0C0}2 \\ \cline{2-4} 
\multicolumn{1}{l|}{\cellcolor[HTML]{C0C0C0}{\color[HTML]{333333} \,$p=0$}} & \multicolumn{1}{l|}{-18}                                                & \multicolumn{1}{l|}{-16}  & \multicolumn{1}{l|}{-6}   \\ \cline{2-4} 
\multicolumn{1}{l|}{\cellcolor[HTML]{C0C0C0}{\color[HTML]{333333}\qquad 1}}     & \multicolumn{1}{l|}{\cellcolor[HTML]{FE0000}{\color[HTML]{333333} -22}} & \multicolumn{1}{l|}{-12}  & \multicolumn{1}{l|}{6}    \\ \cline{2-4} 
\multicolumn{1}{l|}{\cellcolor[HTML]{C0C0C0}{\color[HTML]{333333}\qquad  2}}     & \multicolumn{1}{l|}{-18}                                                & \multicolumn{1}{l|}{0}    & \multicolumn{1}{l|}{26}   \\ \cline{2-4} 
\multicolumn{1}{l|}{\cellcolor[HTML]{C0C0C0}{\color[HTML]{333333}\qquad  3}}     & \multicolumn{1}{l|}{-6}                                                 & \multicolumn{1}{l|}{20}   & \multicolumn{1}{l|}{54}   \\ \cline{2-4} 
\multicolumn{1}{l|}{\cellcolor[HTML]{C0C0C0}{\color[HTML]{333333}\qquad  4}}     & \multicolumn{1}{l|}{14}                                                 & \multicolumn{1}{l|}{48}   & \multicolumn{1}{l|}{90}   \\ \hline
\end{tabular}
}
\caption{Various values of the effective potential with $\tilde{m}=11$. This follows trajectory 3 as shown in figure \ref{NumFig}. The minimum is shown in red.}
\label{NumsMids}
\end{table}

\bibliographystyle{JHEP}
\bibliography{flavorBrokeN}

\end{document}